\documentclass{ws-rv961x669}
\usepackage{ws-rv-van}     
\usepackage{ws-rv-thm}     
\usepackage{subfigure}     
\begin{document}


\begin{center}    
{{\bf High-Energy Neutrinos from Active Galactic Nuclei\footnote{{\bf To be published in Neutrino Physics and Astrophysics, edited by F. W. Stecker, in Encyclopedia of Cosmology II, edited by G. G. Fazio, World Scientific Publishing Company, Singapore, 2022.}}}}
\end{center} 
     
\vspace{1cm}

\author[K. Murase and F.W. Stecker]
{Kohta Murase\footnote{murase@psu.edu}}
\address{Dept. of Physics and Dept. of Astronomy and Astrophysics, Institute for Gravitation and the Cosmos,\\ 
The Pennsylvania State University, University Park, Pennsylvania, USA\\
School of Natural Sciences, Institute for Advanced Study, Princeton, New Jersey, USA,
and Center for Gravitational Physics and Quantum Information, Yukawa Institute for Theoretical Physics, Kyoto, Kyoto, Japan\\ }

and

\author{Floyd W. Stecker\footnote{Floyd.W.Stecker@nasa.gov}}
\address{Astrophysics Science Division, NASA Goddard Space Flight Center,\\
Greenbelt, Maryland, USA\\
and Dept. of Physics and Astronomy, University of California at Los Angeles,\\
Los Angeles, California, USA}


\vspace{1.5cm}

\begin{abstract}
Active Galactic Nuclei (AGN) are sources of high-energy $\gamma$-rays and are considered to be promising candidates to be sources of high-energy cosmic rays and neutrinos as well. We present and discuss various models for ion acceleration and their interactions with matter and radiation leading to high-energy neutrino production.
We consider neutrino production mechanisms in both jet-loud and jet-quiet AGN, focusing on disks and coronae in the vicinity of the central black hole, jet regions, and magnetized environments surrounding the AGN. The IceCube Collaboration has reported high-energy neutrino events that may come from both the jet-loud AGN TXS 0506+056 and the jet-quiet AGN NGC 1068. We discuss the implications of these observations themselves as well as the the origins of the all-sky neutrino intensity.
\end{abstract}

\markboth{High-Energy Neutrinos from Active Galactic Nuclei}{Murase and Stecker} 

\body

\newpage

\tableofcontents

\newpage

\section{Introduction}\label{ra_sec1}
\begin{figure}[t]
\begin{center}
\includegraphics[width=0.8\linewidth]{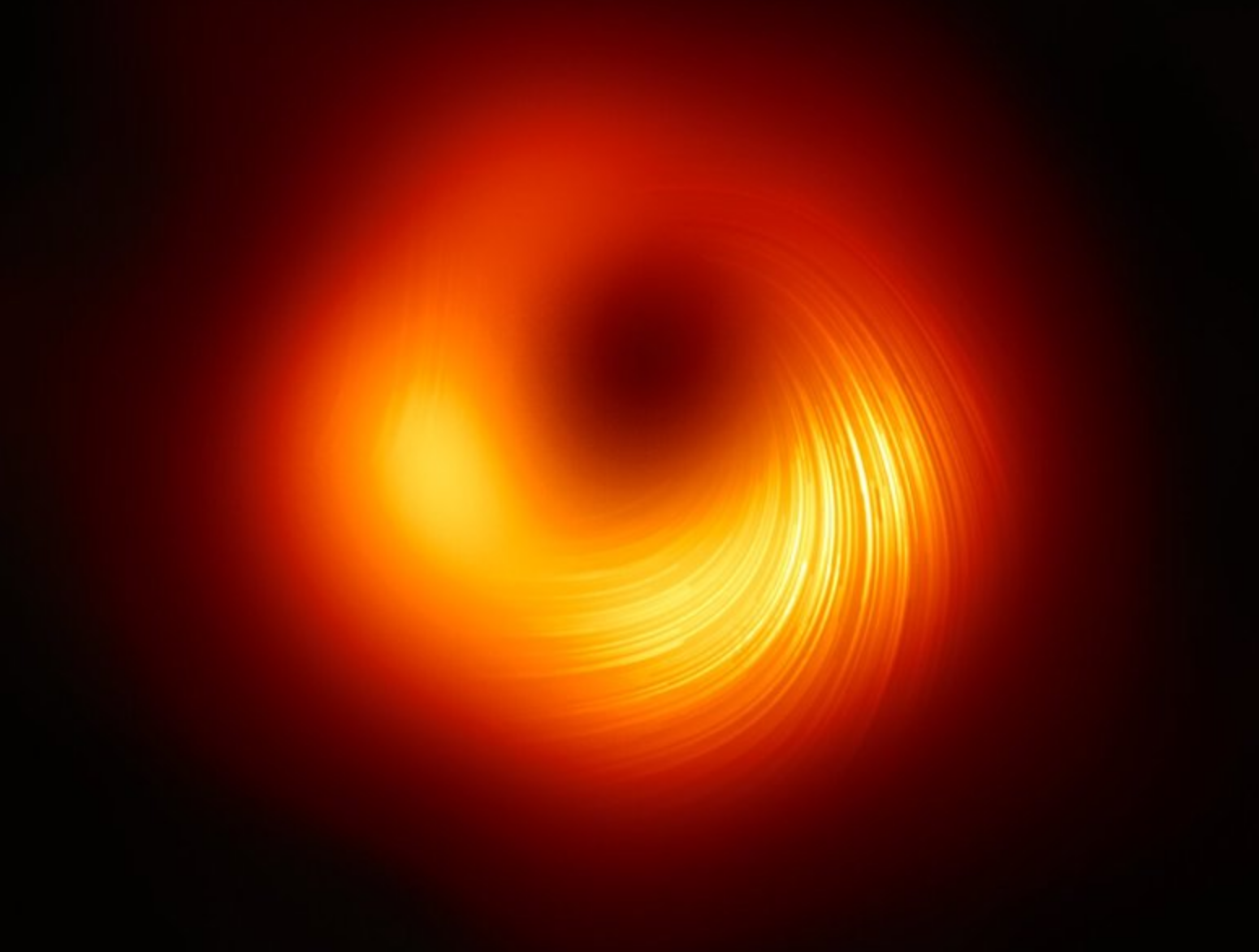}
\caption{Event Horizon Telescope picture of the plasma region around the SMBH in M87 showing polarization field lines with the total intensity underlined~\cite{EventHorizonTelescope:2021bee}.
} 
\label{fig:bh}
\end{center}
\end{figure}

It has been less than six decades since Martin Schmidt first discovered a highly luminous star-like object (quasi-star) at high
redshifts\footnote{They are indicated by clear redshifted Balmer lines.}-- viz., the quasar 3C 273~\cite{Schmidt:1963wkp}. 
Schmidt stated that ``...the explanation in terms of an extragalactic origin seems most direct and least objectionable." The intense luminosity of such objects, quasars and other active galaxies, implied that only gravitational energy could power them. Thus, in the following decade
a scenario emerged whereupon supermassive black holes (SMBHs) at the cores of these objects were invoked to power them.
They are presently known as {\it active galactic nuclei} or AGN. We now have conclusive proof of this SMBH scenario (See Figure~\ref{fig:bh}.)

Black holes are found throughout the Universe. Stellar mass black holes are believed to be the remnants of the explosions of stars with masses greater than $\sim15-20~M_{\odot}$. Intermediate mass black holes and perhaps primordial black holes may also exist.
The subject of this chapter is the SMBHs with masses $M \gtrsim10^5 M_{\odot}$ that reside in the centers of galaxies. 

SMBHs with masses $\gtrsim10^9 M_{\odot}$ have been found at redshifts $z>6$, having formed $\sim 900$~Myr after the Big Bang. Such a fast time scale presents a challenge to theories of the SMBH formation. The new James Webb Space Telescope will be able to make better observations of such high redshift AGN. 

A significant fraction ($\sim1-10$\%) of galaxies are AGN (see, e.g., references~\refcite{2012agn..book.....B,2015ARA&A..53..365N}). Figure~\ref{fig:bh} shows a reconstructed picture of radio emission from the plasma surrounding the SMBH in the nearby AGN M87 that was obtained by the Event Horizon Telescope Collaboration~\cite{EventHorizonTelescope:2021bee}. 

\begin{figure}[t]
\begin{center}
\includegraphics[width=0.8\linewidth]{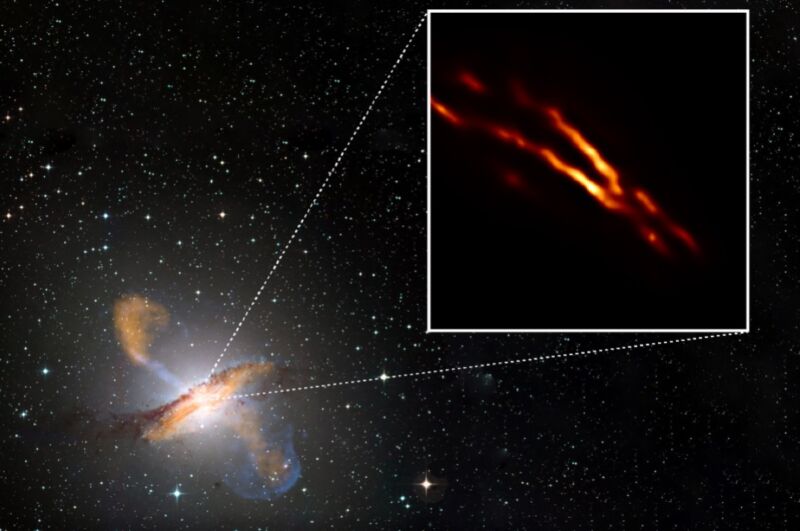}
\caption{Optical and radio images of emission from the jet of the jet-loudAGN Cen A. 
Credit: Radboud University; ESO/WFI; MPIfR/ESO/APEX/A. Weiss et al.~\cite{Weiss:2008wq}; NASA/CXC/CfA/R. Kraft et al.~\cite{Kra+01}; EHT/M. Janssen et al.~\cite{EHTCenA}. 
}
\label{fig:CenA}
\end{center}
\end{figure}

Not all, but some AGN posses powerful jets, clearly indicating that relativistic, collimated plasma flows are being produced by their central SMBHs~\cite{Blandford:2018iot}. Their radiation is observed at distances from the SMBH ranging from $\sim10^{14}~{\rm cm}$ to $\sim{10}^{24}~{\rm cm}$. Figure~\ref{fig:CenA} shows a synthesized image of multiwavelength emission from the nearby jet-loudAGN, Cen A~\cite{EHTCenA}. It is also believed that AGN play important roles in the co-evolution of galaxies and SMBHs~\cite{Fab12,Kormendy:2013dxa}.  

AGN are among the most powerful emitters of radiation in the known universe, emitting a spectrum of electromagnetic radiation ranging from radio wavelengths to high-energy $\gamma$-rays. They are fueled by the gravitational energy of the matter falling onto the SMBH at the center of the AGN, although mechanisms responsible for their efficient conversion of gravitational energy into radiation are not completely understood. They are observed at different wavelengths, and their bolometric luminosity and spectral energy distributions (SEDs) consist of various components including a disk, a corona, a jet, a toroidal region of molecular gas and dust, and a broadline region (BLR). 

%

\begin{figure}[t]
\begin{center}
\includegraphics[width=0.8\linewidth]{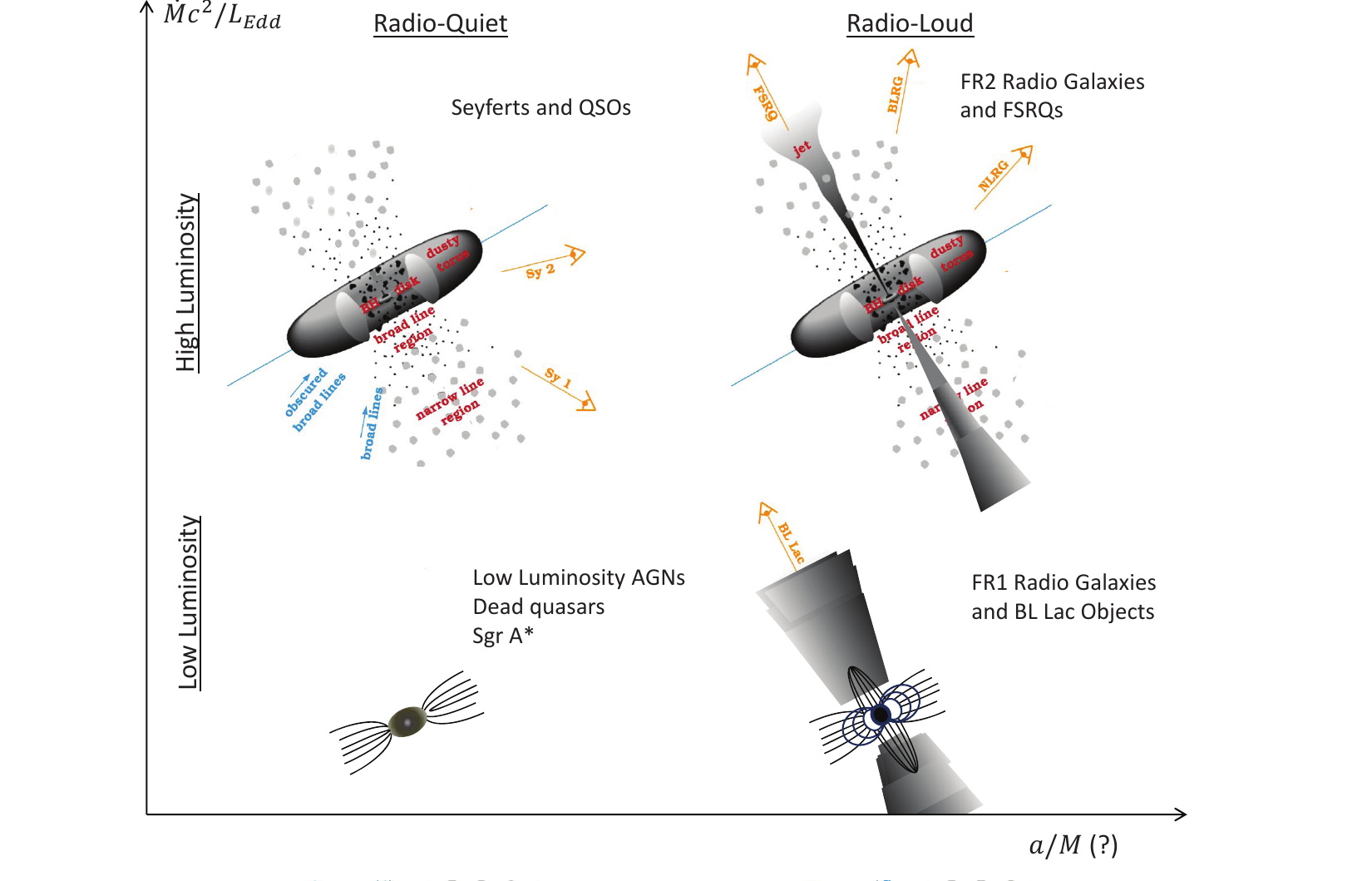}
\caption{Schematic picture of the AGN unification scheme. Adapted from Dermer and Giebels~\cite{Dermer:2016jmw}.} 
\label{fig:AGNunification}
\end{center}
\end{figure}

It is commonly believed that the radio loudness originates from powerful jet activities. Traditionally, radio-quiet (RQ) AGN and radio-loud (RL) AGN represent jet-quiet AGN and jet-loudAGN, respectively (see, e.g., Reference~\cite{Dermer:2016jmw} and Figure~\ref{fig:AGNunification}). 
In general, the AGN classification is pretty diverse depending on wavelengths, and one always has to be careful when the observational classification is connected to physical entities~\cite{Padovani:2017zpf}. Despite various caveats, it is still useful to see the correspondence, and some relevant classes are summarized in Table~\ref{table:AGN}.  

Seyfert galaxies and quasars (or quasi-stellar objects; QSOs) are luminous and classified based on optical observations. Most of AGN are RQ AGN and do not have powerful jets, and only $\sim1-10$\% of them are radio loud, depending on observed wavelengths. This is usually attributed to a change in the accretion mode, the disk magnetization, and the black hole spin. Luminous AGN such as Seyfert galaxies and quasars are believed to be associated with geometrically-thin, radiatively efficient disks~\cite{Shakura:1972te}, which emit most of their accretion power in the optical and ultraviolet bands. 

AGN with lower luminosities (typically less than $10^{42}~{\rm erg}~{\rm s}^{-1}$ in X-rays) are often called low-luminosity (LL) AGN~\cite{Ho:2008rf}. Some LL AGN have Seyfert-like optical spectra but they are more abundant than luminous Seyfert galaxies and quasars. One of the most likely scenarios for low-ionization low-ionization nuclear emission-line regions (LINERs) is a LL AGN. It is believed that LL AGN originate from {\it radiatively inefficient accretion flows} (RIAFs)~\cite{Yuan:2014gma}, which emit only a small fraction of their accretion power as radiation. 

\begin{table*}[bt]
\center{Table 1: Typical classes of AGN observed in the optical band, where $L_{\rm bol}$ is the bolometric luminosity.} 
\begin{center}
\scalebox{0.8}{
\begin{tabular}{c|c|c|c}
\hline Name & $\rho$ [${\rm Mpc}^{-3}$] & $L_{\rm bol}$ [${\rm erg}~{\rm s}^{-1}$] & Comments\\
\hline Low-Luminosity AGN (LL AGN) & $\sim{10}^{-2.5}$ & $\lesssim{10}^{42}$ & weak, low-ionization lines \\
\hline Seyfert galaxy & $\sim{10}^{-3.5}$ & $\gtrsim{10}^{42}$ & strong, high-ionization lines \\
\hline Quasi-Stellar Object (QSO) & $\sim{10}^{-6.5}$ & $\gtrsim{10}^{45}$ & outshining the host galaxy\\
\hline
\end{tabular}
}
\end{center}
\label{table:AGN}
\end{table*}

AGN are the most powerful emitters of high-energy radiation in the known universe. Given their potential for accelerating ions (protons and nuclei) to highly relativistic energies, they have long been considered as potential sites for the production of high-energy cosmic rays and ultrahigh-energy cosmic rays (UHECRs)~\cite{1964ocr..book.....G,Hillas:1984ijl}. 
The detection of high-energy neutrinos from AGN can provide us with clues to the physics in the vicinity of SMBHs, radiation mechanisms from the jets and the possible origin of UHECRs. Hadronic interactions of relativistic nuclei with matter and radiation lead to the production of mesons (mostly pions)~\cite{Stecker:1978ah}. Charged pions generate neutrinos via decay processes, e.g., $\pi^+\to\mu^+\nu_\mu$ followed by $\mu^+\to e^+\nu_e\bar{\nu}_\mu$. Ideas of neutrino production in AGN and early calculations date back to the late seventies and early nineties~\cite{Ber77,Eichler:1979yy,Stecker:1991vm,MSB92,Mannheim:1993jg}, respectively.  

Observational high-energy neutrino astrophysics was born with the discovery of high-energy neutrinos in IceCube~\cite{Aartsen:2013bka,Aartsen:2013jdh}. The energy budget of high-energy neutrinos turned out to be comparable to those of UHECRs and $\gamma$-rays~\cite{Murase:2018utn}.  
Also, observations of neutrinos associated with the AGN designated as TXS 0506+056 could be the first indication of neutrino production in AGN although their physical association is still under debate~\cite{Aartsen2018blazar1,Aartsen2018blazar2,Murase:2018iyl}. Another possible association with a Seyfert galaxy, NGC 1068, was also reported~\cite{IceCube:2019cia,IceCube:2022der}.  
These observations have opened up the prospect of a {\it multimessenger} picture of AGN involving both photons and neutrinos.   

Our understanding of AGN has been improved thanks to the progress of multiwavelength observations as well as theoretical studies including numerical simulations on black-hole accretion and particle acceleration processes. 
In addition, observations of high-energy neutrinos from two AGN, viz., TXS 0506+056 and NGC 1068, have indicated neutrino fluxes larger than those of $\gamma$-rays. 
This may provide a clue as to the observed abundance of extragalactic neutrinos with dim $\gamma$-ray counterparts~\cite{Murase:2015xka}. These have enabled us to have new insights into high-energy neutrino and $\gamma$-ray production in AGN, involving disk-coronae and RIAFs, which can now be confronted with current and near-future multimessenger (photon, neutrino, and cosmic-ray) observations. 

This chapter is organized as follows. 
In Section~\ref{sec:general} we outline the conditions for modeling neutrino production in AGN. 
In Section~\ref{sec:vicinity} we review cosmic-ray acceleration and neutrino production in the vicinity of SMBHs. 
In Section~\ref{sec:jet} we consider neutrino production in inner jets. 
In Section~\ref{sec:environment} we consider models for neutrino production in magnetized environments surrounding AGN. 
In Section~\ref{sec:tidal} we briefly discuss tidal disruption events (TDEs) in the vicinity of SMBHs.
In Section~\ref{sec:summary} we summarize our discussion of high-energy neutrino production in AGN.

\section{General Considerations}\label{sec:general}
\subsection{Conditions for High-Energy Neutrino Production}
\label{sec:model}
There are four conditions for the significant production of high-energy neutrinos in a source.

\begin{itemize} \setlength\itemsep{1em} 
\vspace{12pt}
\item (i) Acceleration of ions (protons and  nuclei) to sufficiently high energies. 
Possible acceleration mechanisms are shock acceleration, magnetic reconnections, and stochastic acceleration in plasma turbulence.

\item (ii) The acceleration rate must overcome the energy loss rates. See equations (\ref{eq:pp}) and (\ref{eq:tpg}) below.

\item (iii) Target media, i.e., matter and radiation, of significant density.

\item (iv) Conditions (i) and (ii), lead to the production of charged mesons, in particular pions, which eventually decay into neutrinos, charged leptons and $\gamma$-rays.
\end{itemize}
\vspace{12pt}
The proton energy loss rate due to inelastic proton-proton interactions is given by
\begin{equation}
t^{-1}_{pp}=n_N\kappa_{pp}\sigma_{pp}c,
\label{eq:pp}
\end{equation}
where $n_N$ is the nucleon density, $\sigma_{pp}$ is the $pp$ cross section, and $\kappa_{pp}\sim0.5$ is the proton inelasticity. The threshold proton energy for pion production is $\varepsilon_p^{\rm th}\simeq1.23$~GeV. 

The proton energy loss rate due to photomeson production is given by~\cite{Stecker:1968uc}
\begin{equation}
t^{-1}_{p\gamma}(\varepsilon _{p})=\frac{c}{2{\gamma}^{2}_{p}} 
\int_{\bar{\varepsilon}_{\rm th}}^{\infty} \! \! \! d\bar{\varepsilon} \, 
{\sigma}_{p\gamma}(\bar{\varepsilon}) {\kappa}_{p\gamma}(\bar{\varepsilon})
\bar{\varepsilon} \int_{\bar{\varepsilon}/2{\gamma}_{p}}^{\infty} 
\! \! \! \! \! \! \! \! \! d \varepsilon \, {\varepsilon}^{-2} 
n_{\varepsilon}, 
\label{eq:tpg}
\end{equation}
where $\varepsilon_p$ is the proton energy in the SMBH rest frame, $\bar{\varepsilon}$ is the photon energy in the proton rest frame, $\sigma_{p\gamma}$ is the $p\gamma$ cross section, and $\kappa_{p\gamma}\sim0.2$ is the proton inelasticity. The threshold condition for pion
production from $p\gamma$ interactions is given by~\cite{Stecker:1968uc}
\begin{equation}
s = m_p^2c^4 + 2m_{p}c^2\bar{\varepsilon}_{\rm th} = (m_{p} + m_{\pi})^2c^4,
\label{cms}
\end{equation}
where $s$ is the Lorentz invariant square of the center-of-momentum energy of the interaction~\cite{Stecker:1968uc} and we have $\bar{\varepsilon}_{\rm th}\simeq0.145$~GeV. 

The photomeson production process near the threshold is dominated by the intermediate production of the $\Delta$ resonance followed by the two-body decay of the $\Delta$ particle which, for neutrino production involves, $p + \gamma \rightarrow \Delta^+ \rightarrow n + \pi^+$. The kinematics of the decay gives a proton inelasticity~\cite{Stecker:1968uc}
\begin{equation}
\kappa_{p\gamma}=1-\left(\frac{\langle\varepsilon_{pf} \rangle}{\varepsilon_{p}}\right) = \frac{1}{2}\left(1-\frac{m_p^2 c^4 - m_{\pi}^2 c^4}{s}\right), 
\end{equation}
with the decay products of the $\Delta^+$ particle, giving $\langle \varepsilon_\pi\rangle \sim \varepsilon_p/5$ which is equivalent to the proton inelasticity~\cite{Stecker:1968uc,Stecker:1978ah}. The four light particles resulting from the decay $\pi^+ \to \mu^+ \nu_\mu \to e^+ \nu_e  \overline \nu_\mu \nu_\mu$ (and the charge-conjugate process) share of energy of the pion roughly equally, giving an average neutrino energy, $\langle \varepsilon_\nu \rangle \sim \langle \varepsilon_\pi\rangle/4 \sim \varepsilon_p/20$.~\cite{Stecker:1978ah}. 

In either $pp$ or $p\gamma$ process, the efficiency of neutrino production can be characterized by the effective optical depth, which is 
\begin{equation}
f_{pp/p\gamma}\equiv \frac{t_*}{t_{pp/p{\gamma}}},
\end{equation}
where $t_*$ is the characteristic time scale specific to models, which can be the proton escape time or the duration of photon emission. 

A comparison between the roughly isotropic extragalactic neutrino flux observed by IceCube and the $\gamma$-ray background flux observed by {\it Fermi},
implies that the fifth condition may be added to the other four conditions given above. Given the large observed neutrino-to-$\gamma$-ray ratio~\cite{Murase:2015xka}, we may add:
\begin{itemize}
\item (v) A high enough target density for intrinsic $\gamma$-ray absorption in a source.
\end{itemize}

If this condition is applied, intra-source electromagnetic cascades will occur, and hadronically produced $\gamma$-rays will then apppear at lower energies after the regeneration. 

\subsection{All-Sky Neutrino Intensity and Multimessenger Connections}
AGN are responsible for various kinds of high-energy phenomena. 
Plasma that accretes onto a SMBH is heated throughout an accretion disk. X-ray emission is produced from hot coronal regions surrounding the SMBH. X-ray emitting AGN are found throughout the extragalactic X-ray sky. Understanding them is also critical to understanding the nonthermal aspects of the Universe. jet-loudAGN are among the most powerful particle acceleration sites in space, and it has been established that they are the dominant sources in the extragalactic $\gamma$-ray background (EGB), especially in the sub-TeV range~\cite{Ajello:2015mfa,TheFermi-LAT:2015ykq,Lisanti:2016jub} (see also Section 4.2). 
Observed emission throughout the electromagnetic spectrum indicates that electrons are accelerated over various distance scales, ranging from the immediate vicinity of SMBHs to kpc scale radio lobes. It is natural that not just electrons but also ions are accelerated by electromagnetic processes. AGN have thus been considered to be promising candidate sources of UHECRs whose origin is a long-standing mystery~\cite{AlvesBatista:2019tlv}. 
As discussed in the previous section, AGN are also promising high-energy neutrino emitters, and they can significantly contribute to the extragalactic neutrino background and the all-sky neutrino intensity given that the Galactic neutrino flux~\cite{Stecker:1978ah,Ahlers:2013xia,Fang:2021ylv} is much smaller than the extragalactic neutrino flux (see Chapter 4).   

In general, the all-sky neutrino intensity, or extragalactic neutrino background, is given by the redshift-weighted line-of-sight integral over redshift~\cite{Stecker:1971ivh,Weinberg:1972kfs,Murase:2015xka} 
\begin{equation}\label{eq:Phinu}
E_{\nu}^2\Phi_{\nu}=\frac{c}{4\pi}\!\int\!\frac{{\rm d}z}{(1+z)^2H(z)}[\varepsilon_{\nu}Q_{\varepsilon_\nu}(z)]\big|_{\varepsilon_\nu=(1+z)E_\nu}\,,
\end{equation}
in units of energy per area, time, and solid angle. Here $E_\nu=\varepsilon_\nu/(1+z)$ is the neutrino energy at $z=0$, $\varepsilon_\nu$ is the neutrino energy in the source frame, $H(z)$ is the redshift-dependent Hubble parameter, and $\varepsilon_{\nu}Q_{\varepsilon_\nu}(z)$ is the neutrino energy generation rate density per logarithmic energy at $z$. For a given cosmic-ray generation rate density, $\varepsilon_{p}Q_{\varepsilon_p}$, one obtains
\begin{equation}\label{eq:Qnu}
\varepsilon_{\nu}Q_{\varepsilon_\nu}\approx\frac{3K}{4(1+K)}{\rm min}[1,f_{pp/p\gamma}]f_{\rm sup}\varepsilon_pQ_{\varepsilon_p}\,,
\end{equation}
where $K$ denotes the ratio of charged to neutral pions with $K\simeq1$ for $p\gamma$ and $K\simeq2$ for $pp$ interactions, and $f_{\rm sup}(\leq1)$ is the suppression factor due to various cooling processes of protons, mesons and muons. 
Evaluating equation~(\ref{eq:Phinu}) leads to the numerical expression
\begin{multline}
E_{\nu}^2\Phi_{\nu}\simeq0.76\times{10}^{-7} ~ {\rm GeV}~{\rm cm}^{-2}~{\rm s}^{-1}~{\rm sr}^{-1}~\\
\times{\rm min}[1,f_{p\gamma}]f_{\rm sup}\left(\frac{\xi_z}{3}\right)\left(\frac{\varepsilon_pQ_{\varepsilon_p}}{{10}^{44} ~ {\rm erg}~{\rm Mpc}^{-3} ~ {\rm yr}^{-1}}\right)\,,
\end{multline}
where $\xi_z$ is a factor accounting for redshift evolution of the neutrino luminosity density~\cite{Waxman:1998yy}. 
For example, $\xi_z\sim0.7$ for the $\gamma$-ray luminosity density evolution of BL Lac objects (BL Lacs), $\xi_z\sim8$ for that of flat-spectrum radio quasars (FSRQs), and $\xi_z\sim3$ for the X-ray luminosity density evolution of AGNs~\citep{Ajello:2013lka,Ajello:2011zi,Ueda:2014tma}. If one uses the number density evolution, we have $\xi_z\sim0.2$ for all BL Lacs and $\xi_z\sim0.1$ for high-synchrotron peak objects, respectively~\citep{Ajello:2013lka}. 

There are two important {\it model-independent} conclusions obtained by the recent multimessenger analyses. 
First, the high-energy neutrino energy budget of the Universe is comparable to those of high-energy $\gamma$-rays and UHECRs~\cite{Murase:2013rfa,Katz:2013ooa,Murase:2018utn}. In general, measured particle intensities can be used for evaluating their energy generation densities by taking into account all energy losses during the cosmic propagation.  
Within order-of-magnitude uncertainties, one gets  
\begin{equation}
\varepsilon_{\nu}Q_{\varepsilon_\nu}|_{0.1~\rm PeV} \sim \varepsilon_{\gamma}Q_{\varepsilon_\gamma}|_{\rm 0.1~\rm TeV} \sim \varepsilon_{p}Q_{\varepsilon_p}|_{10~\rm EeV} \sim 3\times(10^{43}-10^{44})~{\rm erg}~{\rm Mpc}^{-3}~{\rm yr}^{-1}.  
\end{equation}
This suggests that all three messenger particles may have a physical connection. 

Second, the neutrino data at ``medium'' energies (i.e., in the $1-100$~TeV range) suggests the existence of hidden neutrino sources in the sense that the sources are opaque for GeV-TeV $\gamma$-rays. The latest IceCube data showed that the all-sky neutrino intensity is $E_\nu^2\Phi_\nu\sim10^{-7}~{\rm GeV}~{\rm cm}^{-2}~{\rm s}^{-1}~{\rm sr}^{-1}$ at the medium energies, which is higher than the Waxman-Bahcall bound\cite{Waxman:1998yy,Bahcall:1999yr}.
Murase et al.~\cite{Murase:2015xka} showed that $\gamma$-rays associated with the all-sky neutrino flux are inconsistent with the non-blazar component of the EGB measured by {\it Fermi} if the sources are $\gamma$-ray transparent. This conclusion holds for general classes of sources as long as neutrinos and $\gamma$-rays should be co-produced, which has been confirmed by follow-up analyses~\cite{Capanema:2020rjj,Capanema:2020oet}. The combination of neutrino, $\gamma$-ray, and cosmic-ray data require hidden sources if neutrinos are produced via the photomeson production process~\cite{Murase:2015xka}. This is because photons necessary for neutrino production simultaneously prevent high-energy $\gamma$-rays from leaving the sources. The $\gamma\gamma\rightarrow e^+e^-$ optical depth is given by~\cite{Murase:2015xka}
\begin{equation}\label{eq:opticaldepth}
\tau_{\gamma\gamma}(\varepsilon_\gamma^{c})=\frac{\eta_{\gamma\gamma}\sigma_{\gamma\gamma}}{\eta_{p\gamma}\hat{\sigma}_{p\gamma}}f_{p\gamma}(\varepsilon_p)\sim1000 f_{p\gamma},
\end{equation}
where $\eta_{\gamma\gamma}$ is an order-of-unity factor depending on the photon spectrum and $\varepsilon_\gamma^c\approx2m_e^2c^2\varepsilon_p/(m_p\bar{\varepsilon}_\Delta)\sim{\rm GeV}~\left(\varepsilon_\nu/25~{\rm TeV}\right)$.

\section{Disks and Coronae in the Vicinity of Supermassive Black Holes}
\label{sec:vicinity}
Matter in the core region\footnote{The ``core'' is used with different meanings. It means the central parsec region around an AGN in the context of radio observations. In the literature of neutrinos, it indicates a more compact region in the vicinity of a SMBH.} of AGN is manifested by the gravitational potential of SMBHs. 
The radius of the gravitational influence, which is of order the Bondi radius\footnote{The Bondi radius comes from setting escape velocity equal to the sound speed and solving for radius.}, is 
\begin{equation}
R_B \approx \frac{GM_{\rm BH}}{\sigma^2}\simeq5.0\times{10}^{18}~{\rm cm}~\left(\frac{M_{\rm BH}}{10^8~M_\odot}\right){\left(\frac{\sigma}{500~{\rm km}~{\rm s}^{-1}}\right)}^{-2},
\end{equation}
where $\sigma$ is the one-dimensional stellar velocity dispersion. This is a measure of the distance from which gas can accrete onto the SMBH. 

Material accreting onto a SMBH with non-zero angular momenta and at a rate $\dot{m}$ forms an accretion disk, through which a fraction of the gravitational energy is extracted as radiation. 
The radiation luminosity emitted during the accretion from the infinity to the disk radius $R$ is estimated to be 
\begin{equation}
L_{\rm ac}\approx\frac{G\dot{M}M_{\rm BH}}{2R}\simeq3.1\times{10}^{45}~{\rm erg}~{\rm s}^{-1}~{\mathcal R}^{-1}\dot{m}\left(\frac{M_{\rm BH}}{10^8~M_\odot}\right),
\end{equation}
where $\dot M$ is the mass accretion rate, $M_{\rm BH}$ is the black hole mass, and $\mathcal R\equiv R/R_S$ is the normalized radius. Here, $R_S$ is the Schwarzschild radius, 
\begin{equation}
R_S=\frac{2 GM_{\rm BH}}{c^2}\simeq2.8\times{10}^{13}~{\rm cm}~\left(\frac{M_{\rm BH}}{10^8~M_\odot}\right),
\end{equation}
and the normalized accretion rate $\dot m$ is defined as $\dot m\equiv {\dot M}c^2/{L}_{\rm Edd}$, where $L_{\rm Edd}$ is the Eddington luminosity, 
defined as the luminosity of an object that exerts a pressure equal to its gravitational attraction. This is the maximum luminosity beyond which radiation
pressure will overcome the gravitational pull of the object, forcing material away from the object. It is given by
\begin{equation}
L_{\rm Edd}=\frac{4\pi GM_{\rm BH} m_Hc}{\sigma_T}\simeq1.3\times{10}^{46}~{\rm erg}~{\rm s}^{-1}~\left(\frac{M_{\rm BH}}{10^8~M_\odot}\right).
\end{equation}
Luminosities with $L < L_{\rm Edd}$ and $L > L_{\rm Edd}$ are referred to as sub-Eddington and super-Eddington luminosities, respectively.

The inner disk radius is often taken as the radius at the {\it innermost stable circular orbit} (ISCO), which ranges from $GM_{\rm BH}/c^2$ to $9 GM_{\rm BH}/c^2$ depending on the black hole spin. The bolometric luminosity is expressed by,
\begin{equation}
L_{\rm bol}=\eta_{\rm rad}{\dot M}c^2\simeq1.3\times{10}^{45}~{\rm erg}~{\rm s}^{-1}~\eta_{\rm rad,-1}\dot{m}\left(\frac{M_{\rm BH}}{10^8~M_\odot}\right),
\end{equation}
where $\eta_{\rm rad}\sim0.1$ is the radiative efficiency. 
Accretion with $\dot m \gtrsim 10\eta_{\rm rad,-1}$ typically leads to super-Eddington luminosities. 

\begin{figure}[t]
\begin{center}
\includegraphics[width=0.7\linewidth]{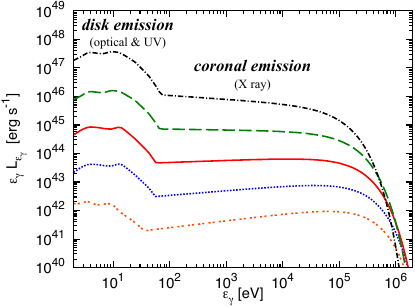}
\caption{SEDs of AGN accretion disks and coronae for different X-ray luminosities from $L_X=10^{41}~{\rm erg}~{\rm s}^{-1}$ to $L_X=10^{45}~{\rm erg}~{\rm s}^{-1}$ (from bottom to top). Adapted from Reference~\cite{Murase:2019vdl}.}
\label{fig:agnsed}
\end{center}
\end{figure}

The radiation is primarily emitted from the accretion disk. Seyfert galaxies and QSOs also show X-ray emission with $L_X\sim(0.01-0.1)L_{\rm bol}$, which is generally interpreted as the Compton upscattering (Comptonization) of disk photons by thermal electrons in hot, magnetized coronae.
The X-ray spectrum is described by a power law with a cutoff, with a spectral slope and cutoff energy that change with the Eddington ratio $\lambda_{\rm Edd}\equiv L_{\rm bol}/L_{\rm Edd}$. See Figure~\ref{fig:agnsed} for spectral energy distributions (SEDs) of emission from disks and coronae. 

Relativistic ions can produce neutrinos via the photomeson production process with disk photons. The big question is whether relativistic particles can be accelerated in such photon-rich environments or not and, if so, how much energy is carried by particles. Observations of a spectral cutoff in X-ray spectra and the absence of 511~keV line emission argue against a cascade origin~\cite{Mad+95,Ricci:2018eir}. However, this does not necessarily mean that particle acceleration does not happen. Recent magnetohydrodynamic (MHD) and particle-in-cell (PIC) simulations have shown that particle acceleration occurs through turbulence and magnetic reconnections.

\subsection{Accretion Disks}
\label{sec:disk}
Quasars and other types of AGN are most powerful continuous emitters of energy in the known universe. These remarkable objects are fueled by the gravitational energy released by matter falling into a SMBH at the galactic center. The infalling matter accumulates in an accretion disk which heats up to temperatures high enough to emit large amounts of ultraviolet and soft X-ray radiation.
In the standard theory of the steady-state accretion~\cite{Pri81}, the disk surface brightness follows from the conservation laws of mass, energy, and angular momentum. A large viscosity is necessary to allow the angular momentum transport, which was introduced by Shakura and Sunyaev~\cite{Shakura:1972te} as the $\alpha$ viscosity parameter. However, its origin and nature had been a mystery for a long time. Now, it is widely accepted that the viscosity originates from MHD turbulence caused by the magnetorotational instability (MRI)~\cite{Balbus:1991ay,Balbus:1998ja}. In a differentially rotating system, angular momentum is transferred from ingoing material to outgoing material through weak magnetic tension, and the ingoing material then further falls toward the center. This is a runway instability, which amplifies magnetic fields and generates strong MHD turbulence. This instability has been confirmed by many numerical simulations and studies.

The fate of accretion flows depends on the accretion rate $\dot{m}$.  
Theoretically, for a sub-Eddington accretion flow with $0.03\alpha_{-1}^2 \lesssim \dot{m} \lesssim10\eta_{\rm rad,-1}^{-1}$, the disk is expected to be geometrically thin and optically thick. The disk spectrum is described by multitemperature black body emission, whose maximum temperature is~\cite{Pri81}
\begin{equation}
T_{\rm disk}=0.488{\left(\frac{3GM\dot{M}}{8\pi R_{\rm ISCO}^3\sigma_{\rm SB}}\right)}^{1/4}\simeq1.7\times{10}^{5}~{\rm K}~{\dot m}^{1/4}{(R_{\rm ISCO}/3R_S)}^{-3/4}R_{S,13.5}^{-1/4},
\end{equation}
where $R_{\rm ISCO}$ is the ISCO radius and $\sigma_{\rm SB}$ is the Stephan-Boltzmann constant. This implies that the typical energy of disk photons is $\varepsilon_{\rm disk}\sim3kT_{\rm disk}\sim10-20$~eV, which typically lies in the UV range. The disk photon spectrum below this peak is $dL_{\rm disk}/d\ln \varepsilon\propto \varepsilon^{4/3}$, so we may approximately write   
\begin{equation}
\varepsilon n^{\rm disk}_\varepsilon=\frac{L_{\rm disk}}{2\pi R^2 \Gamma(4/3) c\varepsilon_{\rm max}} {(\varepsilon/\varepsilon_{\rm max})}^{4/3}e^{-\varepsilon/\varepsilon_{\rm max}},
\end{equation}
where $\Gamma(x)$ is the Gamma function and $\varepsilon_{\rm max}\sim\varepsilon_{\rm disk}$. Note that the spectrum approaches the Rayleigh-Jeans tail at sufficiently low energies. 

\begin{figure}[t]
\begin{center}
\includegraphics[width=0.7\linewidth]{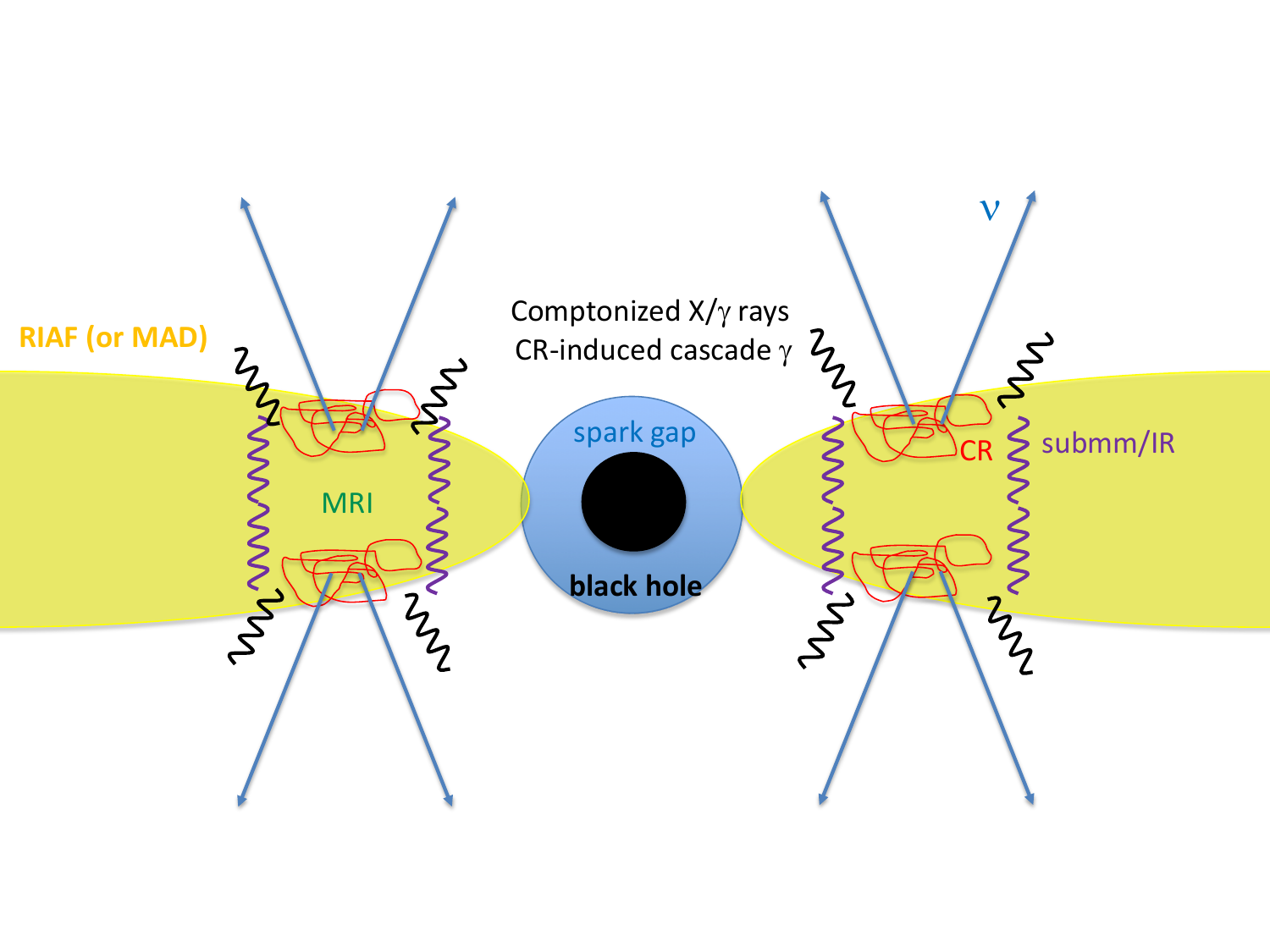}
\caption{Schematic picture of the RIAF/MAD model for high-energy neutrino production. Cosmic rays that are accelerated in the hot disk either through magnetic reconnections or plasma turbulence or both. They interact with gas and radiation from the disk.}
\label{fig:riafmad}
\end{center}
\end{figure}

\begin{figure}[t]
\begin{center}
\includegraphics[width=0.7\linewidth]{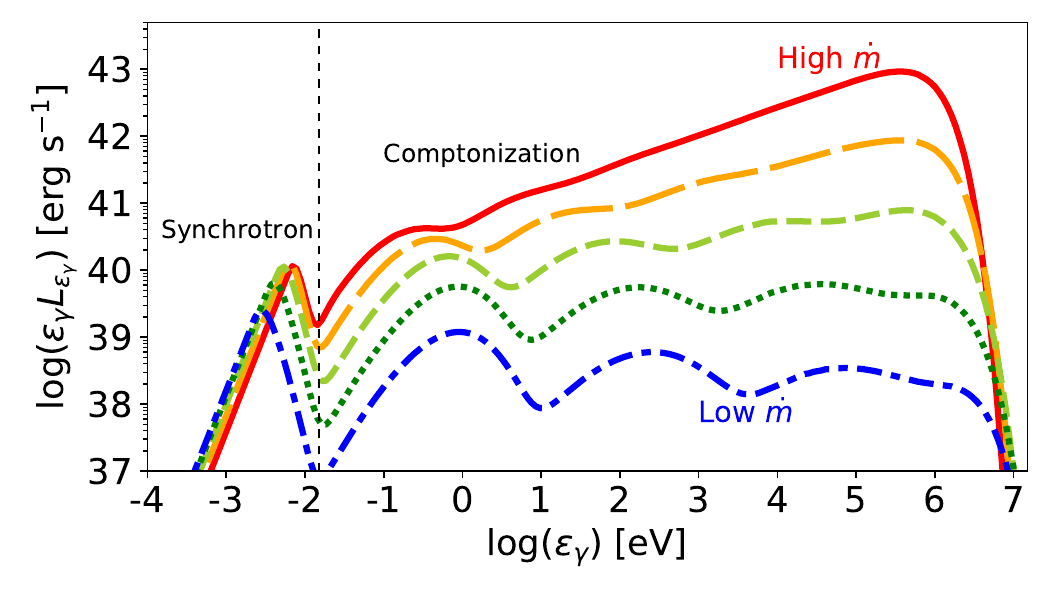}
\caption{SEDs of the RIAF/MAD model for high-energy neutrino production. Cosmic rays that are accelerated in the disk interact with gas and radiation from the disk. Adapted from Kimura et al.~\cite{Kimura:2020thg}.}
\label{fig:riafsed}
\end{center}
\end{figure}


Within radiation-dominated disks, including a super-Edddington ones, Coulomb collision time scales are shorter than the gas infall time scale, so particles are readily thermalized and efficient particle acceleration is not expected. However, the situation is different in RIAFs that are expected to realize in LL AGN. It has been shown that the Coulomb collision time scales for ions is longer than the infall time scale~\cite{TK85,Mahadevan:1997zq}, in which ions can be accelerated by not only magnetic reconnections~\cite{Bednarek:1998jq,Khiali:2015tfa,Ball:2018icx} but also the MHD turbulence~\cite{Dermer:1995ju,Kimura:2014jba,Kimura:2016fjx,Kimura:2018clk,Zhdankin:2018lhq,Comisso:2019frj,Wong:2019dog}. Electrons may also be accelerated via magnetic reconnections~\cite{Ball:2018icx}. 
The existence of the MRI ensures that the disk is magnetized and magnetic reconnections happen, and recent numerical simulations have shown that the RIAFs are indeed the promising sites for particle acceleration~\cite{Hoshino:2013pza,Ball:2016mjx,Ball:2017bpa}. 
Disk properties also depend on the magnetization that may be governed by external magnetic fields carried by the accreting material. Magnetically arrested disks (MADs) may hold back inflowing gas by their strong magnetic fields attached to the disk and generate more powerful jets. Magnetic reconnections in MADs may also accelerate particles to high energies~\cite{Singh:2014jma,Kimura:2020srn}. 
Acceleration by magnetic reconnections may also be accompanied by stochastic acceleration from plasma turbulence~\cite{Zhdankin:2018lhq,Comisso:2019frj,Wong:2019dog}. 
Furthermore, particles may be accelerated in the black hole magnetosphere~\cite{Levinson:2000nx,Hirotani:2016yyk,Hirotani:2017ihv}. When the accretion rate is low enough, the plasma density is so low that a spark gap may form, which has been supported by recent PIC simulations~\cite{Levinson:2018arx,Kisaka:2020lfl}.   

The accretion rate of RIAFs (whether they are MAD or not) is smaller than the critical value, $m_{\rm crit}\approx 0.03\alpha_{-1}^2$. Disks are no longer described by a multitemperature black body spectrum; they are believed to consist of synchrotron radiation and Compton upscattering from thermal electrons (see Figure~\ref{fig:agnsed}). Indeed, RIAFs can successfully explain SEDs of Sgr A$^*$~\cite{Yuan:2003dc,Yoon:2020yew} and LL AGN such as M87~\cite{EventHorizonTelescope:2021dvx}. The number density of photons is typically small, so the photomeson production is important only for LL AGN with an accretion rate close to $\dot{m}_{\rm crit}$. 

On the other hand, relativistic particles should interact with the disk gas during the infall time $t_{\rm fall}\approx R/V_{\rm fall}$, where $V_{\rm fall}\sim\alpha \sqrt{GM_{\rm BH}/R}$, and the effective $pp$ optical depth is estimated to be~\cite{Kimura:2019yjo}
\begin{eqnarray}
f_{pp}\approx n_N (\kappa_{pp}\sigma_{pp})R(c/V_{\rm fall})\approx \frac{8\dot{m}}{\alpha^2}\frac{\kappa_{pp}\sigma_{pp}}{\sigma_T}\sim0.4~\alpha_{-1}^{-2}\dot{m},
\end{eqnarray}
where $n_N$ is the disk nucleon density, $\sigma_{pp}\sim6\times{10}^{-26}~{\rm cm}^2$ is the inelastic $pp$ cross section at PeV energies, $\kappa_{pp}\sim0.5$ is the proton inelasticity, and $\sigma_T$ is Thomson cross section. This estimate suggests that a significant fraction of the energy of the relativistic protons can be depleted for neutrino and gamma-ray production in RIAFs, given that these particles are accelerated by turbulence and magnetic reconnections.

\subsection{Accretion Shock Models}
\label{sec:shock}
\begin{figure}[t]
\begin{center}
\includegraphics[width=0.6\linewidth]{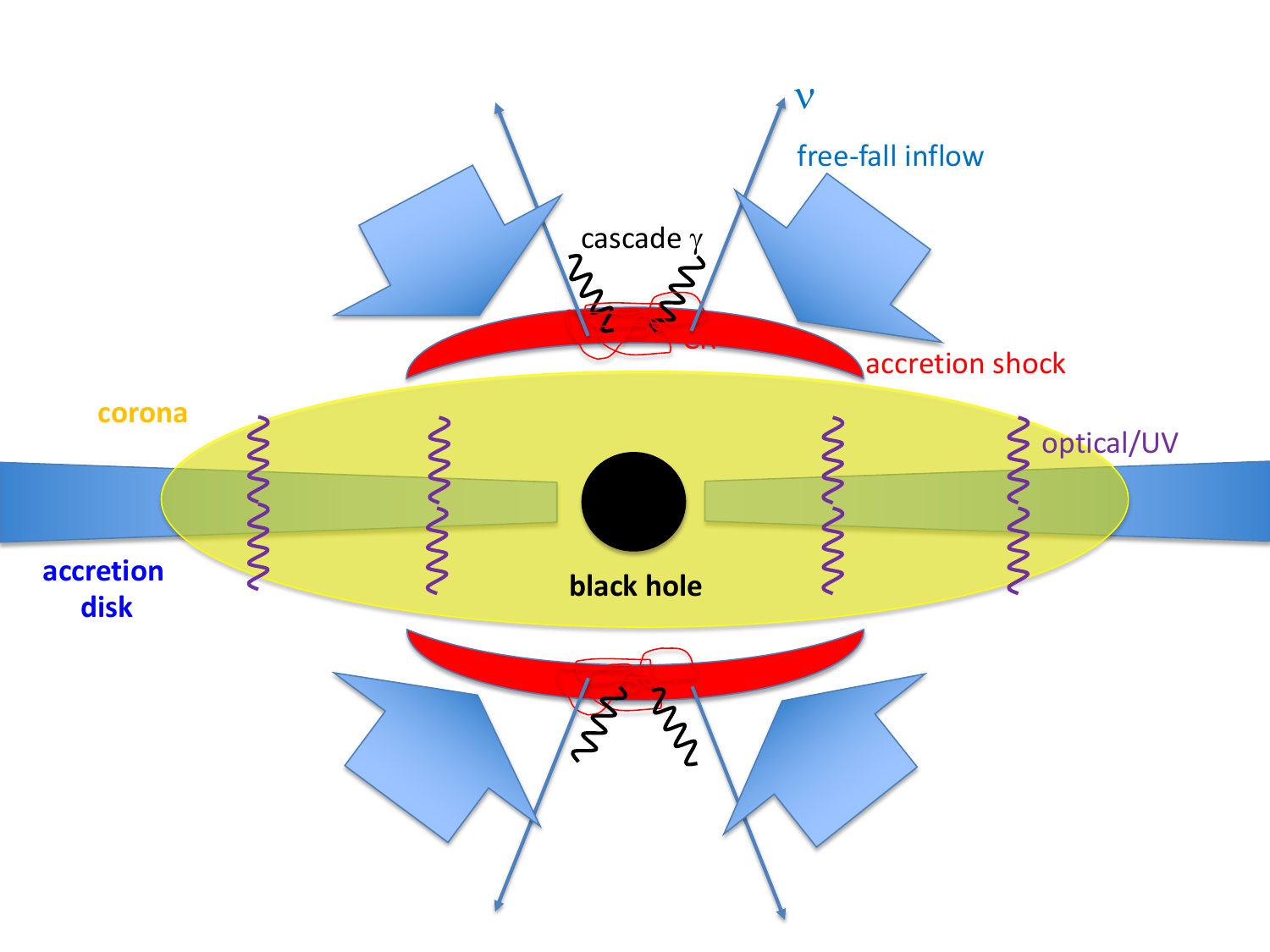}
\caption{Schematic picture of the accretion shock model for high-energy neutrino production. Cosmic rays that are accelerated at possible accretion shocks interact with radiation from the disk.}  
\label{fig:accretionshock}
\end{center}
\end{figure}

The mechanism responsible for the efficient conversion of gravitational energy to observed luminous energy is not yet completely understood. If this conversion occurs partly through the acceleration of particles to relativistic energies~\cite{Stecker:2007zj}, perhaps by shocks formed at the inner edge of the accretion disk~\cite{PK83} (See Figure \ref{fig:accretionshock}.) The interactions of the resulting shock-accelerated high-energy cosmic rays with the intense photon fields produced by the disk or corona surrounding the SMBH, can lead to the copious production of mesons. The photon field in the accretion disk typically peaks in the extreme UV, ${\cal{O}}$(10 eV), known as the "big blue bump"~\cite{Malkan:2020wav,Laor:1989pt}. The coronal radiation is mainly in the X-ray range, and the combined SED is shown in Figure 4 \ref{fig:agnsed}.

The subsequent decay of mesons from interactions between shock accelerated relativistic nucleons and disk and coronal radiation can lead to the generation large fluxes of $\gamma$-rays and high-energy neutrinos. However, since the $\gamma$-rays and high-energy cosmic rays deep in the intense radiation field can lose their energy rapidly and may not leave the source region, these AGN core regions may only be observable as high-energy neutrino sources.

Particle acceleration has been considered in the context of accretion shocks that may be formed by material that almost freely falls onto SMBHs.
The accretion shock velocity is estimated by the free-fall velocity,
\begin{equation}
V_{\rm ff}\approx\sqrt{\frac{GM}{R}}\simeq3.9\times{10}^{9}~{\rm cm}~{\rm s}^{-1}~{\left(\frac{M_{\rm BH}}{10^8~M_\odot}\right)}^{1/2}{\left(\frac{\mathcal R}{30}\right)}^{-1/2}. 
\end{equation}
Note that this velocity is significantly greater than the infall velocity of the accretion flow, which makes the difference in the relative importance of $pp$ and $p\gamma$ interactions, compared to models of magnetically-powered coronae (see the next section). It was proposed that observed X rays originate from electromagnetic cascades induced by relativistic particles~\cite{Stecker:1991vm}, although the existence of a cutoff~\cite{Mad+95,Ricci:2018eir} shows that they are mostly attributed to Comptonized disk photons. 

Theoretically, material should have a non-zero angular momentum and the infall velocity should be smaller than the free-fall velocity. Nevertheless, one cannot exclude the existence of accretion shocks~\cite{Becker:2008uqi} or shocks produced by the Lense-Thirring effect~\cite{Fragile:2008sv} or possible blob collisions~\cite{Alvarez-Muniz:2004xlu}, although efficient dissipation via such shocks has not been manifested in the recent global MHD simulations~\cite{JBSS19,Liska:2018ayk}. If the shock exists, diffusive shock acceleration, which is supported by kinetic simulations, may operate, and the acceleration time scale is 
\begin{equation}
t_{\rm DSA}=\eta_{\rm acc}\frac{\varepsilon_p}{eBc},
\end{equation}
where $\eta_{\rm acc}\sim10 {(c/V_{\rm ff})}^2$ in the Bohm limit. 
Alternatively, particle acceleration by electric fields in a spark gap in the SMBH magnetosphere has been proposed~\cite{Kalashev:2015cma}. However, this mechanism is promising only for LL AGN but it is also unlikely for the standard disk. This is because the plasma density is so high that the quasineutral condition for MHD is usually satisfied~\cite{Levinson:2010fc}. 

For thermal ultraviolet photons in the accretion disk, with $\varepsilon_{\rm disk}\sim10-20~{\rm eV}$, this translates into a characteristic proton energy $\varepsilon_{p}\gtrsim3-10~{\rm PeV}$. The fact that this reaction turns on at such high energies implies that the photons and neutrinos from decaying pions are produced at very high energies too, well above the TeV range. The energy of neutrinos interacting with $\sim10$~eV photons from the accretion disk is expected to be $\varepsilon_\nu \sim1$~PeV. Using $\hat{\sigma}_{p\gamma}\sim0.7\times{10}^{-28}~{\rm cm}^2\sim\kappa_{p\gamma}\sigma_{p\gamma}$ as the attenuation cross section, the effective optical depth is estimated to be
\begin{equation}
f_{p\gamma}\geq n_{\rm disk}\hat{\sigma}_{p\gamma}R\sim50~L_{\rm disk,45.3}{(\mathcal R/30)}^{-1}R_{S,13.5}^{-1}{(10~{\rm eV}/\varepsilon_{\rm disk})},\,\,\,\,\,
\end{equation}
where the lower limit is evaluated when relativistic protons interact with photons during the light crossing time. This result, $f_{p\gamma}\gg1$, implies that cosmic rays are efficiently depleted through the photomeson production. In this sense, the vicinity of SMBHs is ``calorimetric''. Note that the multipion production is dominant at higher energies in the case of thermal photon backgrounds, and $f_{p\gamma}$ cannot decrease with energy. 

In the accretion shock scenario, nonthermal electrons should also be accelerated, and generate $\gamma$-rays via Compton scattering on photons from the accretion disk. The $\gamma$-rays are attenuated or cascaded down to the MeV energy range. The optical depth of the source to $\gamma$-rays from electron-positron pair production is given by
\begin{equation}
\tau_{\gamma\gamma} (\varepsilon_\gamma) \approx \eta_{\gamma\gamma} \sigma_{\gamma \gamma}R n_X {(\varepsilon_p/\tilde{\varepsilon}_{\gamma\gamma-X})}^{\Gamma_X-1} 
\sim 30\frac{L_{X,44}{(\varepsilon_\gamma/\tilde{\varepsilon}_{\gamma\gamma-X})}^{\Gamma_X-1}}{({\mathcal R}/30)R_{S,13.5}(\varepsilon_X/1~{\rm keV})},
\end{equation}      
where $\tilde{\varepsilon}_{\gamma\gamma-X} =m_e^2c^4/\varepsilon_X\sim300~{\rm MeV}~(1~{\rm keV}/\varepsilon_X)$ and $\Gamma_X\sim2$ is the photon index of X-ray emission. This implies that $\gamma$-rays below 10~MeV energies can escape without significant attenuation due to the electron-positron pair production.

\subsection{Magnetically-Powered Coronae}
\label{sec:corona}
%
\begin{figure}[t]
\begin{center}
\includegraphics[width=0.7\linewidth]{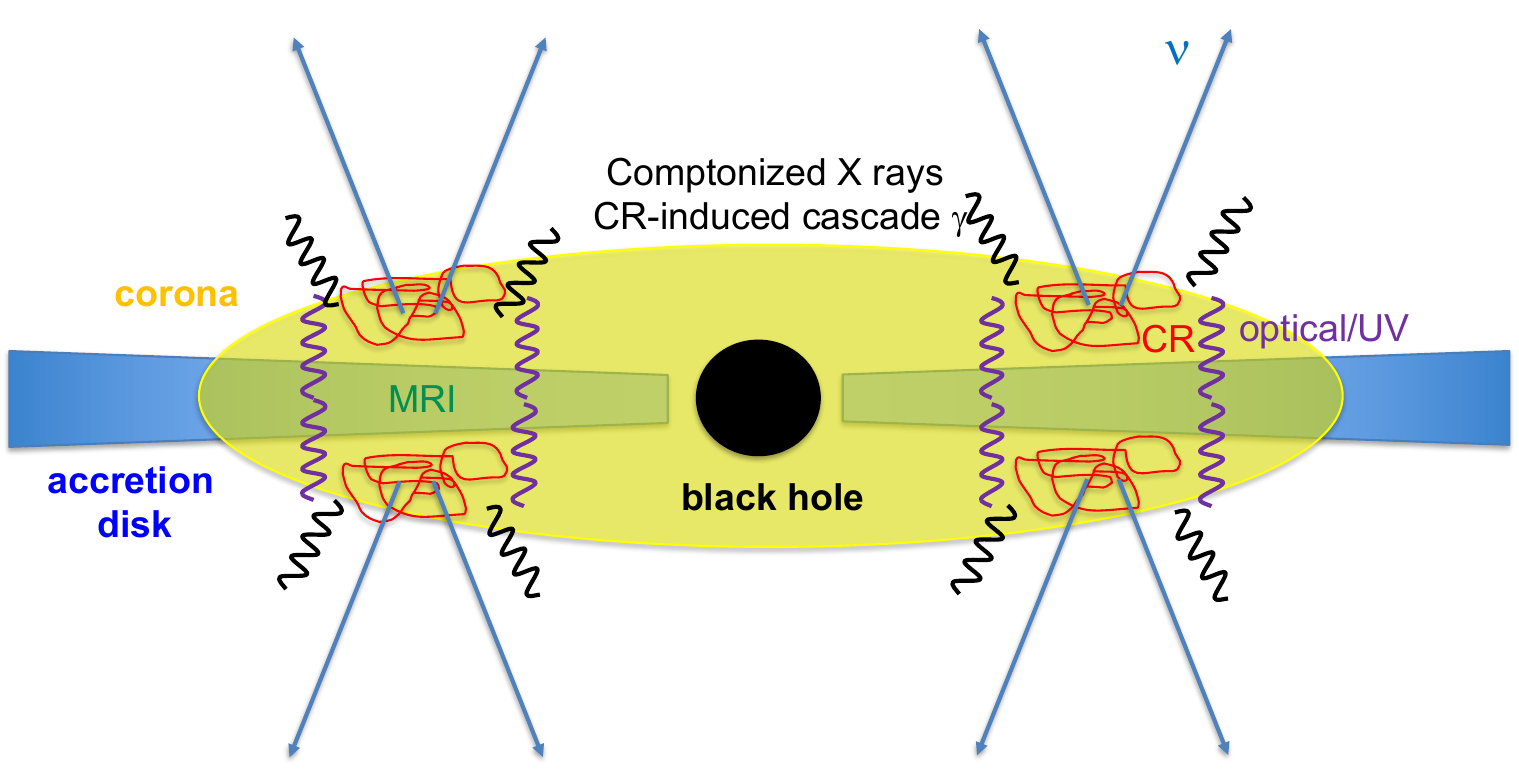}
\caption{Schematic picture of the magnetically-powered corona model for high-energy neutrino production~\cite{Murase:2019vdl}. Cosmic rays that are accelerated in the coronal region interact with coronal plasma, optical and UV photons from the accretion disk, and X-rays from the hot magnetized corona.}
\label{fig:corona}
\end{center}
\end{figure}

In the standard picture of AGN emission, X-ray emission is interpreted as Compton emission by ``thermal'' electrons in high-temperature coronal regions. The X-ray spectrum is described by a power law with a photon index of $\sim 2$ and a high energy cutoff of $\sim10-100$~keV. 
The coronal formation mechanism is still under debate, but the most likely mechanism is motivated by the theory of solar flares. Ordered magnetic fields emerge as a result of the Parker buoyancy instability~\cite{Galeev:1979td}. The stored magnetic energy is then released in the corona via, e.g., magnetic reconnections. In this manner, part of the accretion energy liberated in the disk is transferred to the corona through magnetic fields, heating up the corona during reconnections and eventually radiated away in X-rays~\cite{Merloni:2000gs,Malzac:2001av,Liu:2002ts,Blackman:2009fi}. 
This scenario has been supported by recent MHD simulations~\cite{Miller:1999ix,Jiang:2014wga,JBSS19,Chashkina:2021zox}. By analogy with the Sun, they are formed by some magnetic activity but may be dynamical with high velocities or accompanied by reconnection-induced shocks. 
Note that accretion shocks, which are expected from the free-fall inflow, are different phenomena and have been classified in the context of accretion flow models as discussed above~\cite{Malzac:2001av}.  

X-ray studies suggest that the size of the coronal region is $\sim10-100$ times the Schwarzschild radius of the SMBH (see Figure~\ref{fig:corona}), The coronal plasma is expected to be strongly magnetized, and MHD simulations~\cite{Miller:1999ix,Jiang:2014wga,JBSS19} suggest that the plasma parameter,
\begin{equation}
\beta \equiv \frac{8\pi n_p k T_p}{B^2}
\end{equation}
is low, i.e., $\beta \lesssim 1-3$, where $T_p$ is the proton temperature and $B$ is the coronal magnetic field strength. Modeling of the X-ray emission with the Comptonization mechanism~\cite{Sunyaev:1980fq} implies that the optical depth of the coronal region is 
$\tau_T \sim 0.1-1$~\cite{Merloni:2000gs,Ricci:2018eir}. This allows us to estimate the nucleon density via the relation $n_p \approx \tau_T/(\sigma_TR)$ when the coronal plasma is not dominated by electron-positron pairs. 

As in RIAFs, Coulomb collision time scales are longer than the dissipation time scale in the corona~\cite{Liu:2002ts}. The corona is also strongly magnetized and turbulent, so it is a promising site for particle acceleration. Motivated by this, Murase et al.~\cite{Murase:2019vdl} investigated neutrino and $\gamma$-ray emission in light of the standard magnetically-powered corona model. They primarily considered the stochastic acceleration mechanism, which is slower than diffusive shock acceleration, although fast acceleration may be achieved by magnetic reconnections and termination shocks formed by reconnection-driven outflows~\cite{Chen:2015rcq,Ball:2018icx,Kheirandish:2021wkm}. 

In the magnetically-powered corona model, the Bethe-Heitler pair production process can play a crucial role~\cite{Murase:2019vdl}. This is especially the case if particle acceleration is slower than the shock acceleration in the Bohm limit, by which the maximum energy is limited by the Bethe-Heitler process especially for luminous AGN. This is different from accretion shock models, in which efficient photomeson production with disk photons is dominant. The characteristic energy of protons interacting with disk photons through the Bethe-Heitler process is 
\begin{equation}
\tilde{\varepsilon}_{\rm BH-disk}=0.5m_pc^2\bar{\varepsilon}_{\rm BH}/\varepsilon_{\rm disk} \simeq0.47~{\rm PeV}~{(\varepsilon_{\rm disk}/10~{\rm eV})}^{-1},
\end{equation}
where $\bar{\varepsilon}_{\rm BH}\sim10$~MeV, and note that this energy is below the threshold energy for pion production. Thus, for medium-energy neutrinos in the $10 - 100$~TeV range, the Bethe-Heitler process is more important than the photomeson production, and its effective Bethe-Heitler optical depth is given by~\cite{Murase:2019vdl}
\begin{eqnarray}
f_{\rm BH}&\approx&n_{\rm disk}\hat{\sigma}_{\rm BH}R(c/V_{\rm fall})\sim40~L_{\rm disk,45.3}\alpha_{-1}^{-1}{(\mathcal R/30)}^{-1/2}R_{S,13.5}^{-1}{(10~{\rm eV}/\varepsilon_{\rm disk})},\,\,\,\,\, 
\end{eqnarray}   
where $\hat{\sigma}_{\rm BH}\sim0.8\times{10}^{-30}~{\rm cm}^2$ is the attenuation cross section taking into account the proton inelasticity. 
Even if protons cool mainly via the Bethe-Heitler production process, coronal X rays still provide target photons for the photomeson production, and the effective optical depth is
\begin{eqnarray}
f_{p \gamma}\approx\eta_{p\gamma}\hat{\sigma}_{p\gamma}R(c/V_{\rm fall}) n_X{(\varepsilon_p/\tilde{\varepsilon}_{p\gamma-X})}^{\Gamma_X-1}
\sim2~\frac{\eta_{p\gamma}L_{X,44}{(\varepsilon_p/\tilde{\varepsilon}_{p\gamma-X})}^{\Gamma_X-1}}{\alpha_{-1}({\mathcal R/30)}^{1/2}R_{S,13.5}(\varepsilon_X/1~{\rm keV})},
\end{eqnarray}
where 
\begin{equation}
\tilde{\varepsilon}_{p\gamma-X} = 0.5m_pc^2\bar{\varepsilon}_{\Delta}/\varepsilon_X \simeq 0.14~{\rm PeV}~{(\varepsilon_X/1~{\rm keV})}^{-1},
\end{equation}
$\eta_{p\gamma}\approx2/(1+\Gamma_X)$, and $\bar{\varepsilon}_\Delta \sim0.3$~GeV. Because of $f_{p\gamma} < f_{\rm BH}$, neutrino production is suppressed below the pion production threshold. 

In magnetized coronae, efficient $pp$ interactions can be expected. In this model, the infall time is regarded as the escape time, so the effective $pp$ optical depth is given by
\begin{eqnarray}
f_{pp}\approx n_p (\kappa_{pp}\sigma_{pp})R(c/V_{\rm fall})\sim2~(\tau_T/0.5)\alpha_{-1}^{-1}{(\mathcal R/30)}^{1/2}.
\end{eqnarray}
X-ray observations suggest that coronae of many AGN may be consistent with ion-electron plasma but the moderate pair loading is also possible~\cite{Ricci:2018eir}. The total effective optical depth, $f_{\rm mes}=f_{p\gamma}+f_{pp}$, always exceeds unity, i.e.,
${\rm min}[1,f_{\rm mes}]\sim1$.

Thus, the system is calorimetric and AGN coronae are expected to be efficient neutrino emitters provided that relativistic protons are accelerated in them. The magnetically-powered corona model typically predicts that luminous AGN produce $\sim10-100$~TeV neutrinos rather than $\sim1-10$~PeV neutrinos as in accretion shock models. Also, both $p\gamma$ and $pp$ interactions are important, especially for lower-luminosity objects. In such lower-luminosity objects including LL AGN, $pp$ interactions are more important~\cite{Kimura:2014jba}.

\subsection{All-Sky Neutrino Intensity and MeV $\gamma$-ray Connection}
%
\begin{figure}[t]
\begin{center}
\includegraphics[width=0.6\linewidth]{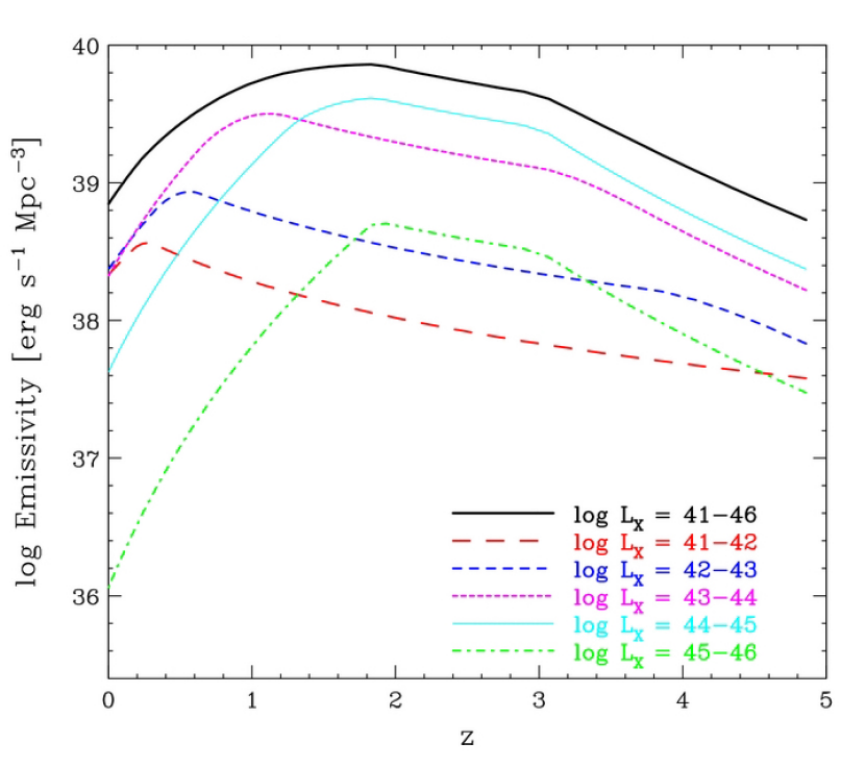}
\caption{X-ray luminosity density of AGN (that are mostly Seyfert galaxies and quasars) as a function of redshift. Adapted from Ueda et al.~\cite{Ueda:2014tma} 
}
\label{fig:agnbudget}
\end{center}
\end{figure}
%
%
\begin{figure}[t]
\begin{center}
\includegraphics[width=0.7\linewidth]{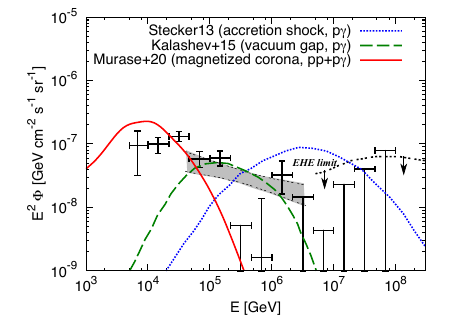}
\caption{Comparison of jet-quiet AGN models~\cite{Stecker:2013fxa,Kalashev:2015cma,Murase:2019vdl} accounting for the all-sky neutrino intensity measured in IceCube. For the IceCube data, the 6 year shower data~\cite{IceCube:2020acn} and 10 year track data~\cite{Stettner:2019tok} are shown as data points and shaded area, respectively, and the extremely high-energy (EHE) limit~\cite{IceCube:2018fhm} is also overlaid.     
}
\label{fig:agnnucore}
\end{center}
\end{figure}
%

RQ AGN that are mostly Seyfert galaxies and quasars are known to be the dominant contributors to the observed X-ray background~\cite{Fabian:1992rt}. 
The differential X-ray luminosity density is given by $L_X^2 d\rho/dL_X$. As shown in Figure~\ref{fig:agnbudget}, the local X-ray luminosity density (at $z=0$) (in the $2-10$~keV band) is 
\begin{equation}
Q_X\equiv\int dL_X \, L_X\frac{d\rho}{dL_X}\sim2\times{10}^{46}~{\rm erg}~{\rm Mpc}^{-3}~{\rm yr}^{-1}. 
\end{equation}
Thus, AGN can largely contribute to the all-sky neutrino intensity if a significant fraction of the thermal energy is carried by cosmic rays. The amount of cosmic rays can be parametrized by the so-called cosmic-ray loading factor,
\begin{equation}
\xi_{\rm cr}\equiv\frac{L_{\rm CR}}{L_{X}},
\end{equation}
which is smaller than unity in the models considered above.  
  
The vicinity of SMBHs perfectly meets the conditions placed by these multimessenger data. In particular, the magnetically-powered corona model gives~\cite{Murase:2019vdl}
\begin{eqnarray}
E_\nu^2\Phi_\nu&\sim&{10}^{-7}~{\rm GeV}~{\rm cm}^{-2}~{\rm s}^{-1}~{\rm sr}^{-1}~\left(\frac{2K}{1+K}\right)\mathcal{R}_{\rm cr}^{-1}\left(\frac{\xi_z}{3}\right)\nonumber\\
&\times&\left(\frac{15f_{\rm mes}}{1+f_{\rm BH}+f_{\rm mes}}\right){\left(\frac{\xi_{\rm CR,-1}L_X\rho_X}{2\times{10}^{46}~{\rm erg}~{\rm Mpc}^{-3}~{\rm yr}^{-1}}\right)}.\,\,\,\,\,\,\,\,
\label{eq:diffuse}
\end{eqnarray}
where ${\mathcal R}_{\rm cr}$ is the conversion factor from bolometric to differential luminosities. Thus, the medium-energy data of the all-sky neutrino intensity in the $10-100$~TeV range and high-energy data above 100~TeV energies may be explained by Seyfert galaxies and quasars. 

The neutrino spectrum shown in Figure~\ref{fig:agnnucore} may indicate a high-energy cutoff. The cutoff, if confirmed, would be useful for constraining some specific models discussed below.
We note that the IceCube Collaboration has reported the detection of a 6.3 PeV neutrino event produced by a Glashow resonance interaction~\cite{IceCube:2021rpz}. Better data with IceCube-Gen2~\cite{IceCube-Gen2:2020qha} are needed for definitive conclusions.

\subsubsection{Some Specific Models}
Stecker et al.~\cite{Stecker:1991vm} provided the first quantitative estimate on the AGN contribution to the all-sky neutrino intensity, using the X-ray luminosity function obtained by the {\it GINGA} satellite. According to the accretion shock scenario, the first-order Fermi acceleration was assumed, in which the maximum energy is around $\varepsilon_{p}^{\rm max}\sim10$~PeV$-10$~EeV. Later, the accretion shock scenario for the X-ray background was excluded, and the dominant fraction of X-rays should originate from thermal electrons. The model was revised based on the MeV $\gamma$-ray background, leading to a factor of 20 reduction~\cite{Stecker:2005hn,Stecker:2013fxa}. 

In the model of Stecker et al.~\cite{Stecker:1991vm} (see also Reference~\cite{Inoue:2019fil}), high-energy neutrinos are produced mainly via $p\gamma$ interactions of shock-accelerated relativistic nuclei~\cite{PK83} interacting overwhelmingly with $\sim10$~eV thermal accretion disk photons from the ``big blue bump". The neutrino spectrum resulting from the decay of the secondary charged pions typically has a peak in the PeV energy range. However, as shown in Figure~\ref{fig:agnnucore}, the IceCube data~\cite{IceCube:2018fhm} may provide a stronger constraint than the MeV $\gamma$-ray constraint. The blue dotted curve violates the EHE limit, so the flux or spectral peak would need to be further lower.

Because of the tension with the IceCube data, Kalashev et al~\cite{Kalashev:2015cma} suggested another model with lower values of the spectral peak in the neutrino spectrum (see Figure~\ref{fig:agnnucore}). They assumed the spark gap scenario, although this assumption would not hold in luminous AGN. The model also only considers photomeson interactions, so the IceCube data above 100~TeV are expected, which may be consistent with the results of the stacking search for infrared-selected AGN~\cite{IceCube:2021pgw}. However, the "medium-energy" neutrino data in the $10-100$~TeV range~\cite{IceCube:2020acn} remain unexplained.  

\begin{figure}[t]
\begin{center}
\includegraphics[width=0.8\linewidth]{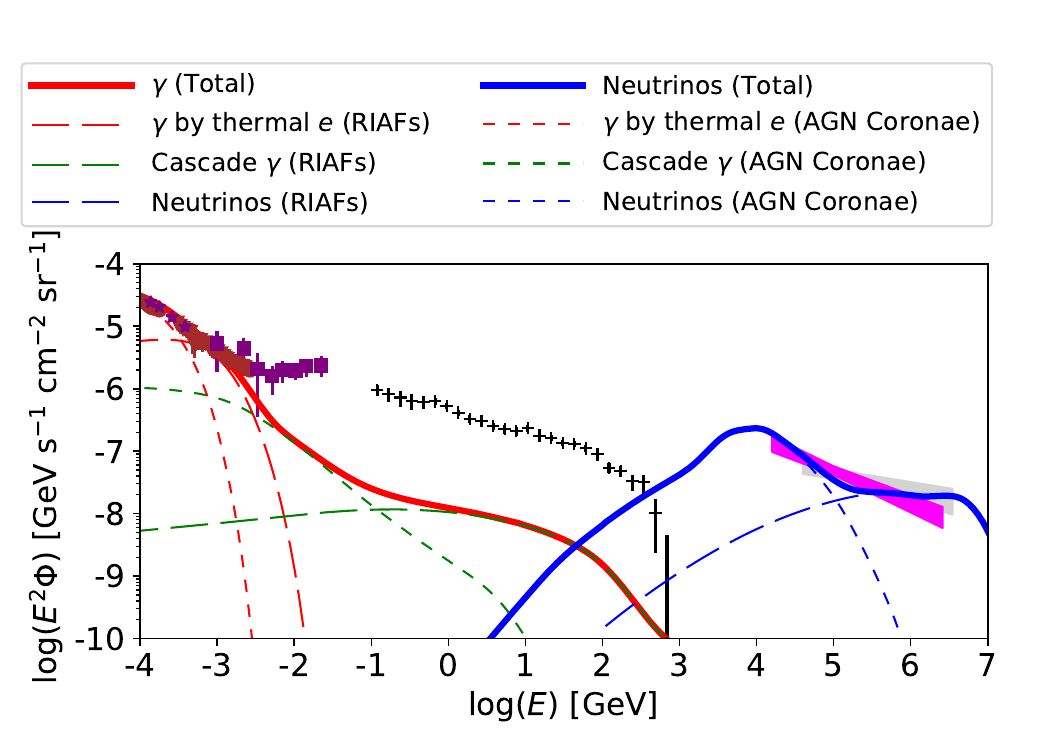}
\caption{AGN scenario to explain the all-sky neutrino intensity from 10~TeV to a few~PeV energies, measured in IceCube, together with the X-ray and MeV $\gamma$-ray backgrounds~\cite{Murase:2019vdl,Kimura:2020thg}. The GeV-TeV $\gamma$-rays may be explained by blazars and radio galaxies, as well as star-forming galaxies (not shown). This provides an example of the multimessenger connection between the TeV-PeV neutrino and the keV-MeV photon backgrounds.
}
\label{fig:agnunification}
\end{center}
\end{figure}

Murase et al.~\cite{Murase:2019vdl} evaluated the AGN contribution to the all-sky neutrino intensity, according to the magnetically-powered corona scenario, using the latest X-ray luminosity function obtained by various X-ray telescopes~\cite{Ueda:2014tma}. They pointed out the importance of the Bethe-Heitler process for neutrino and $\gamma$-ray spectra, which was often ignored in the previous work. If ions are accelerated via an acceleration mechanism that is slower than the diffusive shock acceleration, the maximum energy is also lower, which is predicted to be around $\varepsilon_{p}^{\rm max}\sim1$-10~PeV. Thanks to the Bethe-Heitler suppression and lower maximum energies, neutrinos are typically expected in the $3-30$~TeV range, consistent with the medium-energy neutrino data. Both $pp$ and $p\gamma$ interactions are relevant, the required cosmic-ray luminosity can be more modest for hard cosmic-ray spectra expected in the stochastic acceleration mechanism or magnetic reconnections. 
In this model, the same acceleration mechanism is expected to operate in RIAFs~\cite{Kimura:2014jba,Kimura:2019yjo,Kimura:2020thg}. By applying similar physical parameters to RIAFs for particle acceleration (although the plasma $\beta$ in RIAFs is expected to be larger), higher-energy neutrino data above 100~TeV can simultaneously be explained (see Figure~\ref{fig:agnunification}). In this sense, AGN with different luminosities from LL AGN to luminous AGN including Seyfert galaxies and quasars can explain the all-sky neutrino intensity from a few TeV to a few PeV energies.     

All the models described above generally predict strong connections between neutrinos and MeV $\gamma$-rays. For Seyfert galaxies and quasars, the maximum energy of photons that can escape from the system is around $\sim1-10$~MeV, so electromagnetic energy injected at higher energies should appear in the MeV range. The origin of the MeV $\gamma$-ray background has been unknown, and this neutrino--MeV-$\gamma$-ray connection is intriguing for the future MeV $\gamma$-ray astronomy.  
If neutrinos are produced by $p\gamma$ interactions with disk photons, the proton-induced cascade contribution to the MeV $\gamma$-ray background is minor. On the other hand, in the magnetically-powered corona model with stochastic acceleration, the dominance of the Bethe-Heitler process largely enhances the ratio of $\gamma$-rays to neutrinos. Secondary pairs can also be reaccelerated and energized by turbulence, in which the dominant fraction of the MeV $\gamma$-ray background could be explained. 
Alternatively, primary electrons may be accelerated by magnetic reconnections, in which Seyfert galaxies can account for the MeV $\gamma$-ray background if only a fraction of the thermal energy is used for particle acceleration with a steep nonthermal tail. Such a scenario was suggested by Stecker and Salamon~\cite{Stecker:2001dk} and Inoue et al.~\cite{Inoue:2007tn}. 
More recently, LL AGN have also been suggested as the sources of the MeV $\gamma$-ray background~\cite{Kimura:2020thg}. In this scenario, the MeV $\gamma$-ray background is attributed to Compton emission by thermal electrons in the hot RIAF plasma. To identify the sources and discriminate among different possibilities, both auto-correlation and cross-correlation searches are necessary, which will be feasible with future MeV $\gamma$-ray telescopes~\cite{Inoue:2014ona}. 

%
%

\subsection{NGC 1068 and Detectability of Nearby AGN}
\label{RQ-LL}
\begin{figure}[t]
\begin{center}
\includegraphics[width=0.8\linewidth]{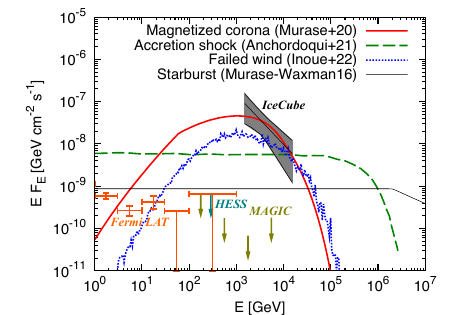}
\caption{All-flavor neutrino and $\gamma$-ray SEDs of NGC 1068. The IceCube neutrino data~\cite{IceCube:2022der}, the {\it Fermi} $\gamma$-ray data, and the MAGIC and HESS upper limits are shown, together with neutrino spectra of three different models~\cite{Murase:2019vdl,Anchordoqui:2021vms,Inoue:2022yak,Murase:2016gly}. 
}
\vspace{-1.\baselineskip}
\label{fig:NGC1068}
\end{center}
\end{figure}

\begin{figure}[t]
\begin{center}
\includegraphics[width=0.6\linewidth]{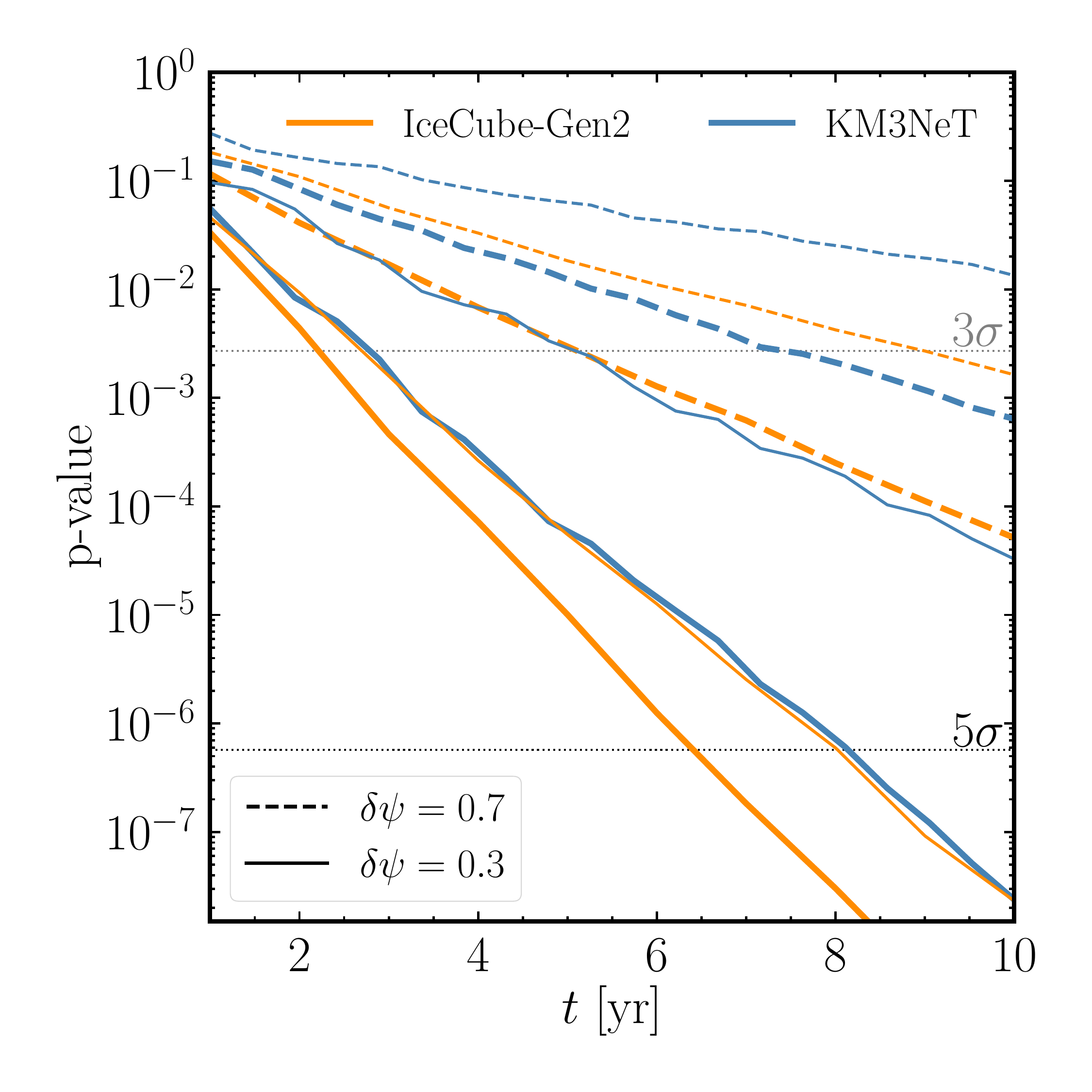}
\caption{
Future prospects for detecting high-energy neutrinos from nearby AGN with stacking analyses using IceCube-Gen2 and KM3Net. The $p$-values as a function of observation time are shown for different angular resolutions~\cite{Kheirandish:2021wkm}. 
\label{fig:nearbyagn}
}
\vspace{-1.\baselineskip}
\end{center}
\end{figure}

The IceCube Collaboration reported a $4.2\sigma$ excess from a detection of 79 neutrinos in the vicinity of NGC 1068 in their 10-year analysis of known $\gamma$-ray sources~\cite{IceCube:2022der}. NGC 1068 is a type 2 Seyfert galaxy, which is also one of the starburst galaxies with AGN, and it has been predicted that NGC 1068 is among the brightest neutrino source in the IceCube sky~\cite{Murase:2016gly}. Contrary to that of M82, the infrared luminosity of NGC 1068 is comparable to the value required by starburst models to explain the all-sky neutrino intensity in the sub-PeV range. 

The neutrino luminosity detected by IceCube at TeV energies is more than an order of magnitude greater than the equivalent $\gamma$-ray luminosity in the 0.1-100 GeV range found by the Fermi-LAT Collaboration~\cite{Fermi-LAT:2019pir,Fermi-LAT:2019yla}.
Thus, although the {\it Fermi} data in the GeV range can be explained to be from pion-decay $\gamma$-rays produced by cosmic rays injected by starbursts and AGN~\cite{Yoast-Hull:2013qfa}, such models violate upper limits found by TeV $\gamma$-ray telescopes. The MAGIC Collaboration reported a search for $\gamma$-ray emission in the very-high-energy band~\cite{MAGIC:2019fvw}. 
No significant $\gamma$-ray signal was detected by MAGIC during 125 hours of observation of NGC 1068. This null result provides a 95\% CL upper limit to the $\gamma$-ray flux above $200~{\rm GeV}$ of $5.1 \times 10^{-13}~({\rm cm^2 \, s})^{-1}$. Thus, the neutrino flux found by the IceCube Collaboration cannot be explained by inelastic $pp$ interactions in the starburst region. In fact, NGC 1068 can be defined to be the epitome of a {\it hidden neutrino source}. As noted in the previous section, such hidden sources have been independently invoked to explain the $10-100$~TeV neutrino data in IceCube~\cite{Murase:2015xka}, and AGN models described above may account for those multimessenger observations.

As an example, the results of the magnetically-powered corona model~\cite{Murase:2019vdl} are shown in Figure~\ref{fig:NGC1068}, where the IceCube data are explained mainly by inelastic $pp$ interactions. We note that $L_X=3\times{10}^{43}~{\rm erg}~{\rm s}^{-1}$ is adopted~\citep{Marinucci:2015fqo}. The required cosmic-ray pressure may be higher than that for the all-sky neutrino intensity, but details depends on parameters. MeV gamma-ray counterparts are predicted, which are good targets for future gamma-ray telescopes such as {\it AMEGO-X} and {\it eASTROGAM}. 
On the other hand, GeV emission should have a different origin such as starburst activities. 

Within the framework of the accretion shock model of Stecker et al.~\cite{Stecker:1991vm}, Inoue et al.~\cite{Inoue:2019yfs} and Anchordoqui et al.~\cite{Anchordoqui:2021vms} calculated the high-energy neutrino flux from NGC 1068. Protons are accelerated up to a maximum energy $\cal{O}$($10^7-10^8)$~GeV with an $E^{-2}$ power-law spectrum resulting from scattering off magnetic field irregularities in a shock at radius $R\sim(10-30)R_S$~\cite{Protheroe:1992qs,Szabo:1994qx}. 
To be consistent with the observed TeV neutrino flux level, $\sim3\times10^{-8}~{\rm GeV}~{\rm cm}^{-2}~{\rm s}^{-1}$ (for all flavors), the neutrino cutoff energy is constrained to $\lesssim 50$~TeV~\cite{Kheirandish:2021wkm}, corresponding to $\varepsilon_p^{\rm max}\lesssim1$~PeV, so the shock acceleration efficiency needs to be less efficient than the Bohm limit.   

All AGN core models shown in Figure~\ref{fig:NGC1068} take account of the observational fact that NGC 1068 is Compton thick~\cite{Marinucci:2015fqo}. Anchordoqui et al.~\cite{Anchordoqui:2021vms} assumed that cosmic rays escape and can interact with a lot of target material to produce both charged and neutral pions via $pp$ collisions. However, such a large column density is typically attributed to disk winds or toroidal regions outside the corona~\cite{Marchesi:2018rgw}, so cosmic rays may not interact so efficiently. 
Murase et al.~\cite{Murase:2019vdl} conservatively considered $pp$ interactions with the material in the coronal region during the infall time (which implies that cosmic rays are rather confined in the plasma). 
Models also considered the photomeson production due to interactions with the X-ray photons surrounding the SMBH and the thermal ultraviolet photons from the inner edge of the accretion disk. More recently, Inoue et al. proposed failed winds as a possible site of diffusive shock acceleration~\cite{Inoue:2022yak}.
 
Both processes result in the production of neutral pions as well as the charged pions. Murase et al.~\cite{Murase:2019vdl} calculated electromagnetic cascades induced by cosmic-ray protons, whereas Inoue et al.~\cite{Inoue:2019yfs} studied the case of dominant electron acceleration, considering only the $\gamma$-ray attenuation without computing electromagnetic cascades. To avoid overshooting the {\it Fermi} data, Inoue et al.~\cite{Inoue:2019yfs} further introduced a screen region, assuming that the coronal region is more compact than the scale of disk photon fields.  

As of this writing the excess of neutrino events found in the direction of NGC 1068 has not reached the discovery level. Further observations are necessary in order to confirm the NGC 1068 excess and to test the AGN models. 
Detections with future neutrino telescopes such as KM3Net~\cite{Adrian-Martinez:2016fdl} and IceCube-Gen2~\cite{IceCube-Gen2:2020qha} would also be more promising. KM3Net has a better sensitivity for AGN in the southern hemisphere with a greater angular resolution. IceCube-Gen2 is expected to be $\sim10$ times bigger in volume. As can be seen in Figure~\ref{fig:nearbyagn}, if AGN are responsible for $10-100$~TeV neutrinos the signal should be detected using stacking analyses, by which we can determine whether or not the AGN core regions are responsible for the all-sky neutrino intensity. Stacking searches are also powerful for LL AGN. 
Kimura et al.~\cite{Kimura:2019yjo,Kimura:2020thg} studied the detectability of nearby LL AGN as the sources of high-energy cosmic neutrinos, and promising targets include NGC 4565, NGC 3516, and NGC 4258.  

Notably, in any of the models described above, neutrino production is efficient and the system is calorimetric (i.e., $f_{pp/p\gamma}\gtrsim1$). One typically expects $L_{\nu}\propto L_X$ given that the cosmic-ray loading factor $\xi_{\rm CR}$ is similar. Using this simple scaling, the effective local number density and corresponding typical X-ray luminosity are estimated to be $\sim3\times{10}^{-6}~{\rm Mpc}^{-3}$ and $\sim10^{44}~{\rm erg}~{\rm s}^{-1}$, respectively, if their all-sky neutrino intensity is $E_\nu^2\Phi_\nu\sim3\times{10}^{-8}~{\rm GeV}~{\rm cm}^{-2}~{\rm s}^{-1}~{\rm sr}^{-1}$. 
Seyfert galaxies and quasars are so abundant that they are not constrained by multiplet or other auto-correlation searches with the current IceCube data~\cite{Murase:2016gly}. 
Nevertheless, there are some chances to find excess neutrino emission from nearby bright objects in near future. X-ray bright AGN are promising, but X-ray emission from AGN is often obscured by the surrounding molecular torus or winds, so the sources that intrinsically luminous are expected to be the most promising. Within the {\it Swift}-BAT BASS catalog, the intrinsically brightest AGN include Circinus Galaxy, ESO 138-G001, NGC 7582, Cen A, NGC 1068, NGC 424, and CGCG 164-019. Interestingly, many of them are located in the southern hemisphere, and the most promising source in the IceCube sky was found to be NGC 1068.  In reality, the detectability of neutrinos also depends on the detector location and zenith angle of sources, and Kheirandish et al.~\cite{Kheirandish:2021wkm} investigated the detectability of individual Seyfert galaxies with the current IceCube and future telescopes such as KM3Net and IceCube-Gen2 (See also Reference~\cite{Kimura:2019yjo}.)

\section{Inner Jets and Blazars}
\label{sec:jet}
Observational properties of AGN vary with viewing angles~\cite{Antonucci:1993sg,Urry:1995mg}. 
A small fraction of AGN have powerful jets and are often RL AGN. 
If the jets are misaligned with the line-of-sight direction, they are called misaligned AGN that are often classified as radio galaxies. The radio galaxies are further divided into Fanaroff-Riley (FR) I and II galaxies. The host galaxies of RL AGN are typically elliptical galaxies, and only $\sim1-10$\% of AGN harbor powerful jets although weak jets may exist even for RQ AGN. FR II galaxies are brighter and more powerful, and their host galaxies are often quasars. Among jet-loud AGN, AGN with powerful jets that point to us are called blazars. In the unification scheme, BL Lacs are regarded as on-axis counterparts of FR I galaxies, whereas quasar-hosted blazars (QHBs) that are predominantly FSRQs correspond to FR-II galaxies viewed on the jet axis. 
One should keep in mind that the observational classification is not always tied to the physical classification. For example, a fraction of Seyfert galaxies have powerful jets that are detected by $\gamma$-rays with {\it Fermi}-LAT.  

\begin{figure}[t]
\begin{center}
\includegraphics[width=0.7\linewidth]{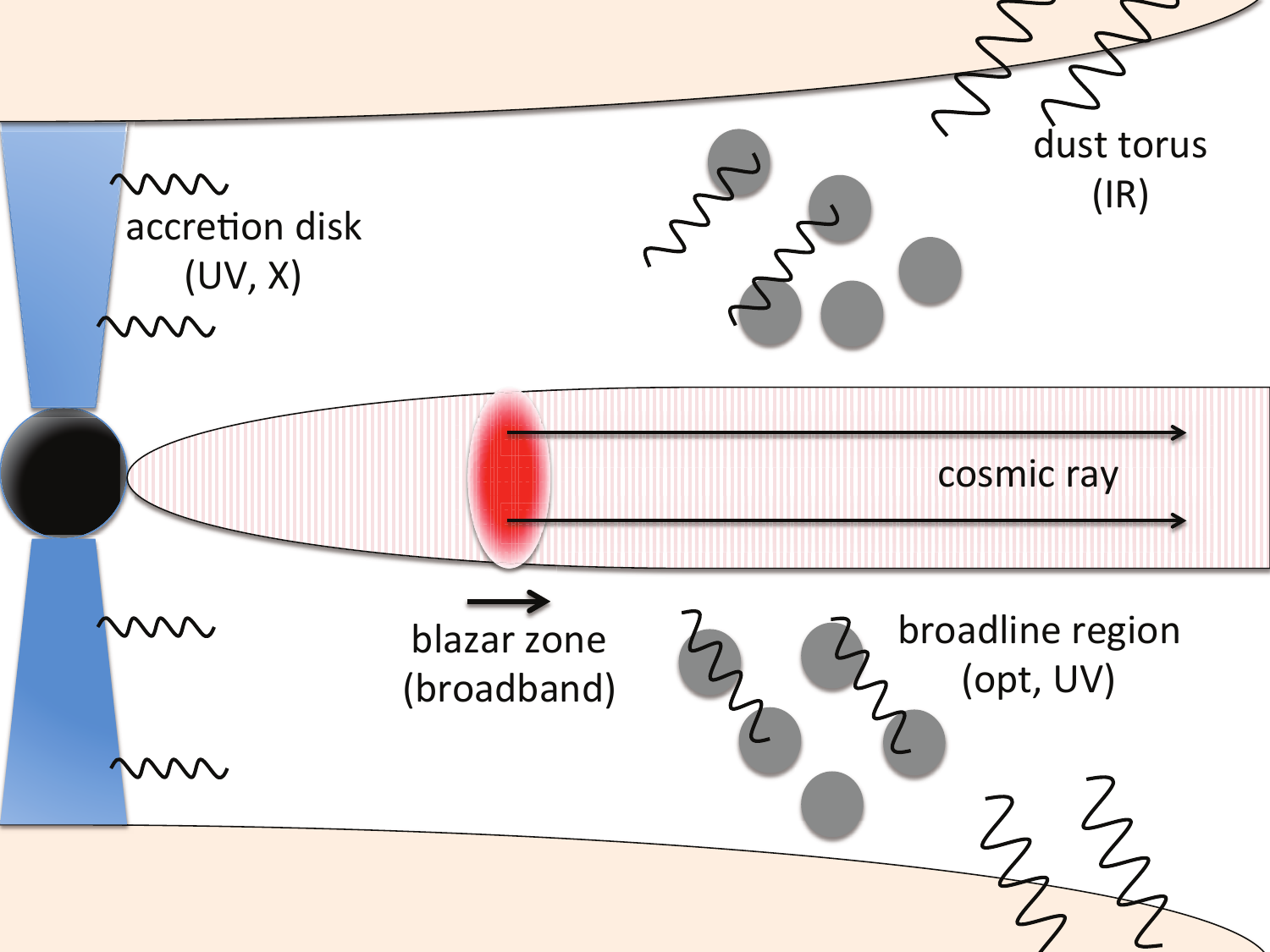}
\caption{Schematic picture of neutrino and $\gamma$-ray production in inner jets of AGN~\cite{Murase:2014foa}. Ions that are accelerated in the jets interact with nonthermal radiation from electrons in the blazar zone, and further interact with scattered disk photons and BLR photons, as well as infrared photons coming from a toroidal region of molecular gas and dust.  
}
\label{fig:blazarschematic}
\vspace{-1.\baselineskip}
\end{center}
\end{figure}

Thanks to the progress of numerical similations, there has been significant progress in our understanding of accretion and jet physics~\cite{Davis:2020wea}.   
The most promising jet production mechanism is the Blandford-Znajek (BZ) mechanism~\cite{Blandford:1977ds}, in which the rotation energy of a SMBH is extracted as the Poyting flux through ordered magnetic fields anchored from the accreting material to the black hole. 
Recent numerical simulations~\cite{McKinney:2006tf,Tchekhovskoy:2011zx} have revealed that the BZ power follows
\begin{equation}
L_{j}=\eta_j\dot{M}c^2\simeq 1.3\times10^{46}~{\rm erg}~{\rm s}^{-1}~\eta_j \dot{m} M_{\rm BH,8}, 
\end{equation}
where $\eta_j$ is the jet efficiency, which can be $\eta_j\sim0.3-1$ for disks are MADs~\cite{McKinney:2012vh}. 

The jet at the base is Poynting dominated, which is also supported by the recent observations and MHD simulations~\cite{Blandford:2018iot}. Observations of radio galaxies such as M87 suggest that the jet is accelerated and collimated within the radius of gravitational influence $R_B$. Jet-loud AGN are observed at different wavelengths, and it is likely that different emission regions contribute to the observed SEDs. The typical blazar emission, which originates from inner jets, is believed to be particle dominated~\cite{Ghisellini:2009fj}, and a significant fraction of the Poynting energy needs to be converted into kinetic energy or particle energy within $R_{B}$. 
The comoving size of a blob, from which blazar emission is radiated, is 
\begin{equation}
l'_b \approx \delta c t_{\rm var}/(1+z)\simeq 3.0\times{10}^{16}~{\rm cm}~\delta_1 t_{\rm var,5} {(1+z)}^{-1},  
\end{equation}
where $\delta$ is the Doppler factor and $t_{\rm var}$ is the variability time scale. For blazars, one expects $\delta \sim \Gamma$, where $\Gamma$ is the bulk Lorentz factor. The emission radius measured from the SMBH is estimated to be
\begin{equation}
r_b \approx \delta l'_b \simeq 3.0\times{10}^{17}~{\rm cm}~\delta_1^2 t_{\rm var,5} {(1+z)}^{-1},
\end{equation}
and the isotropic-equivalent radiation luminosity $L_{\rm rad}$ is related to the comoving radiation luminosity $L'_{\rm rad}$ as
\begin{equation}
L_{\rm rad}=\delta^4 L'_{\rm rad}\sim \frac{\delta^4}{2\Gamma^2}{\mathcal P}_{\rm rad},    
\end{equation}
where ${\mathcal P}_{\rm rad}$ is the total radiation power of the two-sided jets, which should carry a fraction of $L_j$. 

Mechanisms of energy dissipation and resulting particle acceleration in the inner jets have been under debate~\cite{Blandford:2018iot}. 
Traditionally, internal shocks and shock acceleration mechanisms have been considered. Recently, magnetic reconnections and stochastic acceleration are more actively discussed as the primary mechanism of particle acceleration~\cite{Hoshino:2012xm,Sironi:2015eoa,Guo:2015ydj,Werner:2016fxe,PSSG19,Zhang:2021akj}. They may better explain rapid time variability and extremely hard spectra, which are observed in some of blazar flares.

\subsection{Blazars}
\label{sec:blazar}
Blazar SEDs are known to have a two-component, or two-hump, structure (see, e.g., Reference~\cite{Boettcher:2013wxa}). 
The lower-frequency component is interpreted as synchrotron emission radiated by relativistic electrons that are primarily accelerated in jets, while the  higher frequency component is commonly attributed to Compton upscattering off the same population of the primary electrons. This is the so-called leptonic scenario for the $\gamma$-ray origin. 
Theoretically it is natural that not only electrons but also ions are accelerated whether the acceleration mechanism is shock acceleration or magnetic reconnections. If the dominant origin of the high-energy $\gamma$-ray component originates from ions, this is called the hadronic scenario. 
Hadronic scenarios, where observed $\gamma$-rays are attributed to either proton synchrotron emission~\cite{Aharonian:2000pv,Mucke:2000rn} or proton-induced electromagnetic cascades inside~\cite{Mannheim:1993jg} or outside the source~\cite{Essey:2009ju,Essey:2010er,Murase:2011cy}, have primary electrons that are responsible for the low-energy hump, so they also belong to lepto-hadronic models in the sense that both primary ions and electrons are injected in the calculations.  

\begin{figure}[t]
\begin{center}
\includegraphics[width=0.7\linewidth]{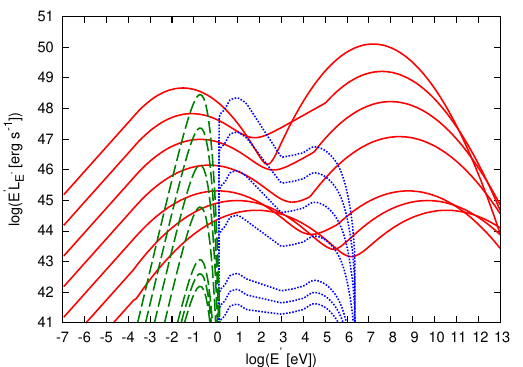}
\caption {Photon SEDs of blazars, where only continuum emission is shown. Solid, dotted, and dashed curves represent the nonthermal jet component, the accretion disk component, and the torus infrared component, respectively. 
From top to bottom, the radio luminosity varies from $\log (L_{\rm 5~GHz}) = 41$ to $\log (L_{\rm 5~GHz})=47$. Adapted from Reference~\cite{Murase:2014foa}.}
\label{fig:blazarsed}
\end{center}
\end{figure}

Characteristic blazar SEDs are shown in Figure~\ref{fig:blazarsed}. 
There is a tendency that blazar SEDs evolve with luminosities, which is called the blazar sequence~\cite{Fossati:1998zn,Ghisellini:2008zp}. The robustness of the blazar sequence is still under debate because of the selection bias~\cite{Padovani:2017zpf}, but it broadly indicates that luminous blazars tend to have lower peak energies. According to the leptonic scenario, QHBs are interpreted as Compton dominated objects, where targets photons can be provided by not only the accretion disk but also the BLR and a torus of molecular gas and dust. Continuum emission from the disk and toroidal region are also shown in Figure~\ref{fig:blazarsed}. 

High-energy neutrinos are produced in lepto-hadronic models, whether the $\gamma$-ray origin is leptonic~\cite{Atoyan:2001ey,Murase:2014foa,Rodrigues:2017fmu} or hadronic~\cite{Mannheim:1993jg,Muecke:2002bi,Essey:2010er,Murase:2011cy,Cerruti:2014iwa,Petropoulou:2015upa,Diltz:2015kha}. 
Synchrotron photons from primary electrons serve natural targets for the photomeson production~\cite{MB89}, and cosmic rays predominantly interact with these photons in jets of BL Lacs. Using parameters of BL Lac objects with the synchrotron radiation luminosity $L_{\rm rad}^s\sim{10}^{45}~{\rm erg/s}~$ and the synchrotron peak energy $\varepsilon_s\sim10$~eV, we have
\begin{eqnarray}
f_{p\gamma}(\varepsilon_p)\simeq7.8\times{10}^{-4}L_{\rm rad,45}^s \Gamma_1^{-4}{t}_{\rm var,5}^{-1}{(\varepsilon_s/10~{\rm eV})}^{-1}
\left\{\begin{array}{ll}
{(\varepsilon_{\nu}/{\varepsilon}_{\nu}^{b})}^{\beta_h-1} 
& \mbox{($\varepsilon_p \leqq {\varepsilon}_{p}^{b}$)}
\\
{(\varepsilon_{\nu}/{\varepsilon}_{\nu}^{b})}^{\beta_l-1} 
& \mbox{(${\varepsilon}_{p}^{b} < \varepsilon_p$)}
\end{array} \right.
\end{eqnarray}
where ${\varepsilon}_{\nu}^{b}$ is the neutrino energy corresponding to the photomeson production with photons at $\varepsilon_s$, $\beta_l\sim1.5$ and $\beta_h\sim2.5$ are the low-energy and high-energy photon indices, respectively.  

For QHBs such as FSRQs, external target photons are important for $r_b<r_{\rm BLR}$, where $r_{\rm BLR}$ is empirically~\cite{DElia:2002ujp,Liu:2006ja,Tavecchio:2008vq} 
\begin{equation}
r_{\rm BLR}\approx{10}^{17}~{\rm cm}~L_{\rm AD,45}^{1/2},
\end{equation}
where $L_{\rm AD}$ is the accretion disk luminosity. The BLR luminosity is related to the disk luminosity through
\begin{equation}
L_{\rm BL}\approx f_{\rm cov}L_{\rm AD},
\end{equation}
where $f_{\rm cov}\sim0.1$ is the covering factor of the BLR. 
The origin of BLRs is under debate, and it may be supplied by clumpy winds from the accretion disk. BLRs are usually seen for luminous AGN but not in LL AGN including FR-I galaxies. 
Given $r_b<r_{\rm BLR}$, the effective optical depth for the photomeson production is~\cite{Murase:2014foa} 
\begin{equation}
f_{p\gamma}\simeq5.4\times{10}^{-2}~f_{\rm cov,-1}L_{\rm AD,46.5}^{1/2},
\end{equation}
which is applied to cosmic rays escaping from the blob. 
The typical energy of BLR photons is in the ultraviolet range, so the resulting energy of neutrinos is $\varepsilon_\nu \gtrsim 1$~PeV. 

Active galaxies typically possess a torus of molecular gas and dust. The radius of this torus is typically of order a parsec and is proportional to the square root of the AGN luminosity. It is given by~\cite{Suganuma:2005ty,Cleary:2006pe,Kishimoto:2010nu,Malmrose:2011ne,Kishimoto:2011hz}
\begin{equation}
r_{\rm DT}\approx 2.5\times{10}^{18}~{\rm cm}~L_{\rm AD,45}^{1/2},
\end{equation}
and the infrared luminosity is estimated to be
\begin{equation}
L_{\rm IR}\approx0.5L_{\rm AD}.
\end{equation}
The effective optical depth for the photomeson production is~\cite{Murase:2014foa}
\begin{equation}
f_{p\gamma}\simeq0.89~L_{\rm AD,46.5}^{1/2}{(T_{\rm IR}/500~{\rm K})}^{-1},
\end{equation}
where $T_{\rm IR}$ is the dust temperature. The resulting energy of neutrinos is $\varepsilon_\nu \gtrsim 100$~PeV. This also implies that UHECRs are efficiently depleted in inner jets of luminous blazars, which is especially the case if UHECRs are heavy nuclei~\cite{Murase:2011cy,Murase:2014foa}. 

%

%
\begin{figure}[t]
\begin{center}
\includegraphics[width=0.7\linewidth]{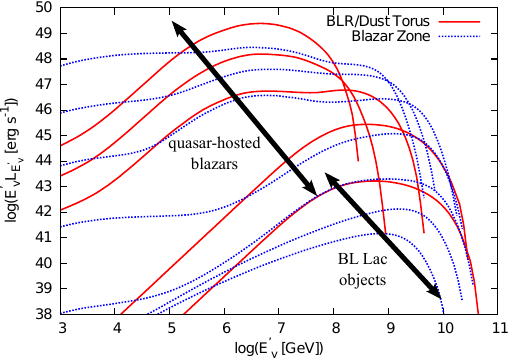}
\caption{Differential luminosity spectra of photohadronic neutrinos from blazars~\cite{Murase:2014foa}, where muon neutrino spectra are shown for the cosmic-ray spectral index $s_{\rm cr}=2.0$, with neutrino mixing being considered. 
From top to bottom, the radio luminosity varies from $\log (L_{\rm 5~GHz})=47$ to $\log (L_{\rm 5~GHz})=41$ as in Figure~\ref{fig:blazarsed}.}
\label{fig:blazarnused}
\end{center}
\end{figure}
%

Whether the blazar sequence is robust or not, one reaches the conclusion that the photomeson production efficiency is higher as the radiation luminosity is larger. Moreover, one may introduce the following phenomenological relationship, 
\begin{equation}
L_{\nu} \propto L_{\rm rad}^{\gamma_{\rm Lw}}, 
\end{equation}
where $\gamma_{\rm Lw}$ is the luminosity-weighting index. 
The cosmic-ray loading factor can be introduced as
\begin{equation}
\xi_{\rm cr}\equiv\frac{L_{\rm cr}}{L_{\rm rad}},    
\end{equation}
where $L_{\rm cr}$ is the isotropic-equivalent cosmic-ray luminosity, and $L_{\rm rad}$ is the bolometric nonthermal luminosity from the jet, which is a fraction of the nonthermal electron luminosity. Given that the acceleration mechanism is common in blazars with different luminosities, it is reasonable to assume that the cosmic-ray loading factor is constant, in which one typically expects $\gamma_{\rm Lw}\sim3/2$ for QHBs and $\gamma_{\rm Lw}\sim2$, respectively~\cite{Murase:2014foa}. On the other hand, if the observed $\gamma$-rays have the hadronic origin, one expects $L_{\nu}\sim L_{\nu}$, corresponding to $\gamma_{\rm Lw}\sim1$~\cite{Petropoulou:2015upa,Padovani:2015mba}.  

Blazar neutrino SEDs corresponding to their photon SEDs in Figure~\ref{fig:blazarsed} are shown in Figure~\ref{fig:blazarnused}. 
The results can be regarded as the neutrino blazar sequence in a sense. However, one should keep in mind that the results for luminous blazars such as QHBs are not sensitive to details of the blazar sequence, as long as external photon fields are dominant as target photons for the photomeson production process~\cite{Murase:2014foa,Rodrigues:2017fmu}. 
The effective optical depth for the photomeson production sharply increases around the pion production threshold for thermal target fields, and it never decreases with energy thanks to the multipion production. Thus, the resulting neutrino spectra cannot be steeper than cosmic-ray spectra. Indeed, as seen in Figure~\ref{fig:blazarnused}, the spectra of blazar neutrinos are predicted to be hard especially below the PeV range.


\subsection{All-Sky Neutrino Intensity and GeV-TeV $\gamma$-Ray Connection}
Blazars are strong $\gamma$-ray emitters and their emission appears to make up most of the EGB in the GeV-TeV range~\cite{Ajello:2015mfa}. 
About 50\% of the EGB can be explained by {\it Fermi}-detected blazars as point sources. The rest is called the diffuse isotropic $\gamma$-ray background (IGRB)~\cite{Fermi-LAT:2014ryh}. 
Analyzing the photon sky map shows that blazars dominate the $\gamma$-ray sky above 50~GeV~\cite{TheFermi-LAT:2015ykq,Lisanti:2016jub}, while the origin of the IGRB remains more uncertain especially below GeV energies. Star-forming galaxies significantly contribute to the EGB below 1-10~GeV energies~\cite{Thompson:2006np,Fields:2010bw,Stecker:2010di,Tamborra:2014xia,Roth:2021lvk,Blanco:2021icw}, although radio galaxies may also give a non-negligible contribution to the IGRB~\cite{Inoue:2014ona,Fornasa:2015qua}.  

\begin{figure}[t]
\begin{center}
\includegraphics[width=0.7\linewidth]{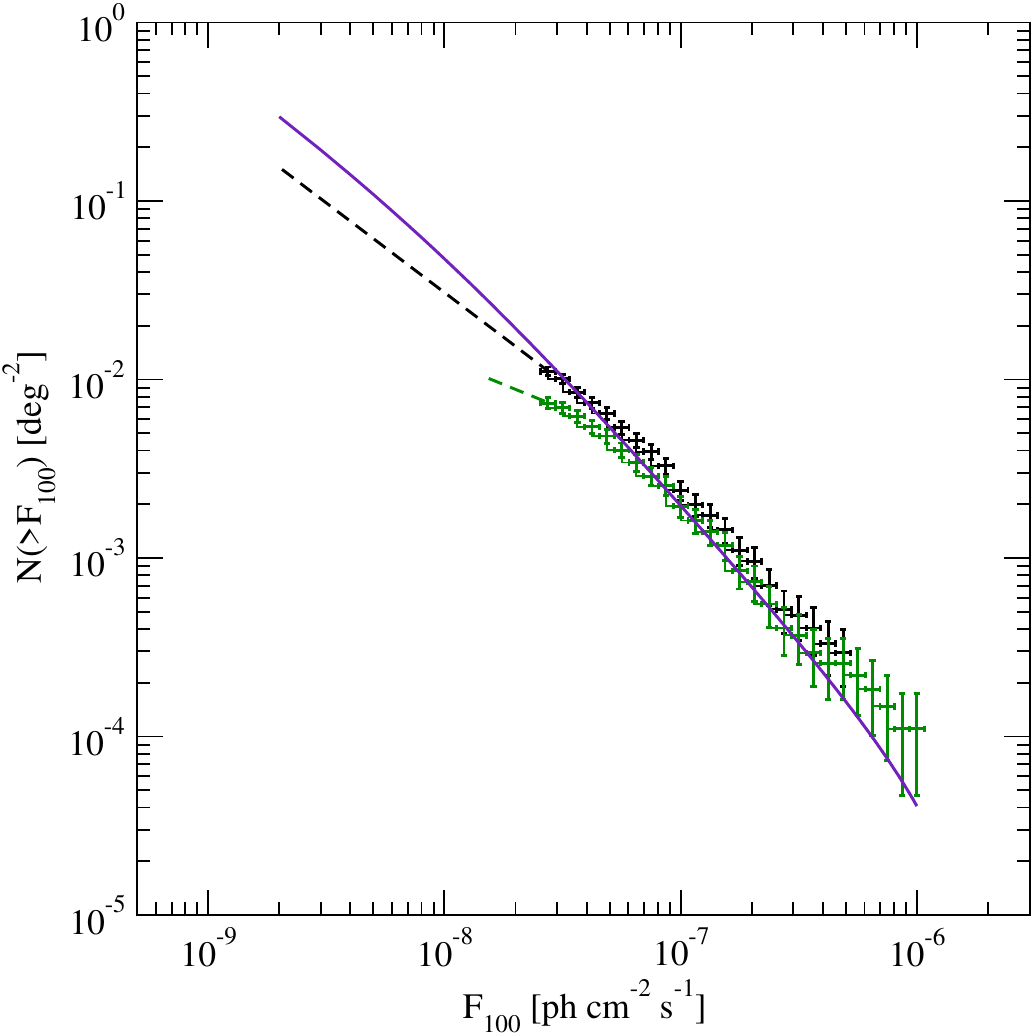}  
\caption{Bright end of the source count distributions for all blazars (including BL Lacs; black data points) and FSRQs (light green data points). The dashed lines are the faint-end slopes determined by including a modeled Monte Carlo {\it Fermi}-LAT efficiency. A fit to the data taking account of faint-end source confusion is shown by the solid (purple) line~\cite{Stecker:2010di}.}
\label{blazarSC}
\end{center}
\end{figure}

\begin{figure}[t]
\begin{center}
\includegraphics[width=0.7\linewidth]{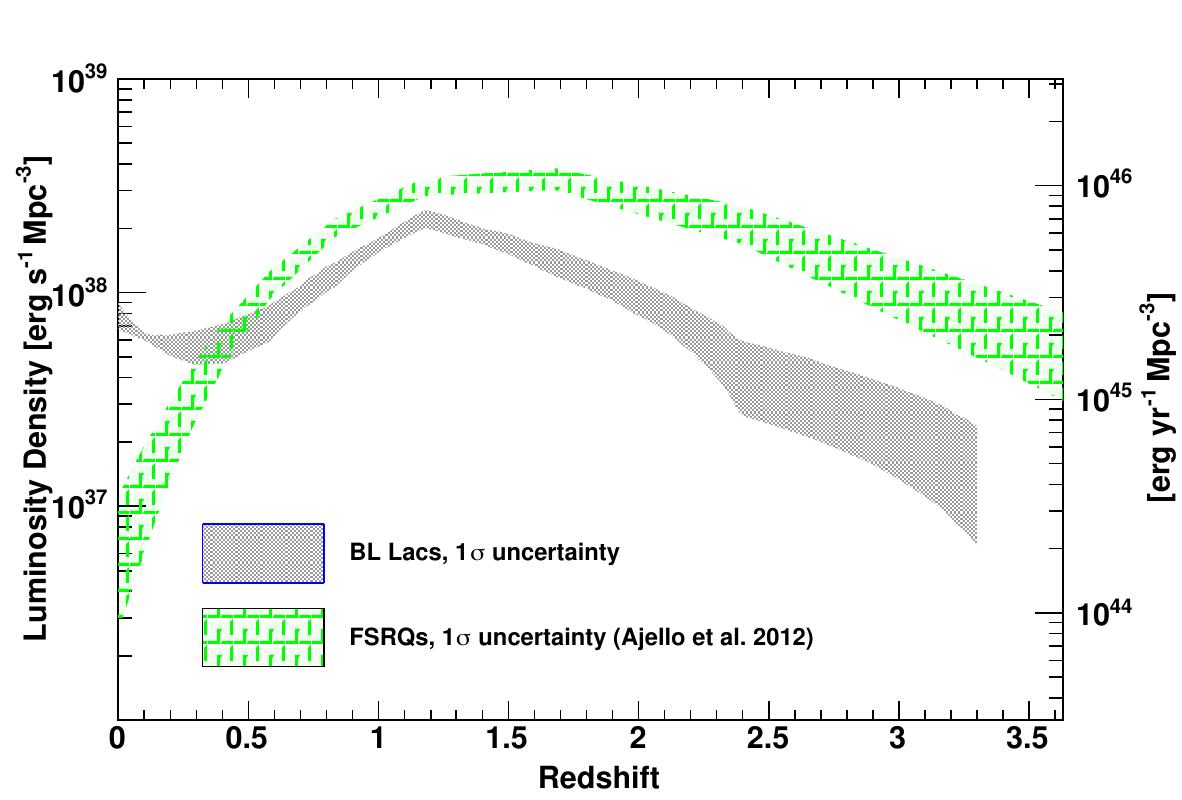}
\caption{Luminosity density of $\gamma$-rays from blazars as a function of redshift~\cite{Ajello:2013lka}. 
}
\label{gammalum}
\end{center}
\end{figure}

\begin{figure}[t]
\begin{center}
\includegraphics[width=0.7\linewidth]{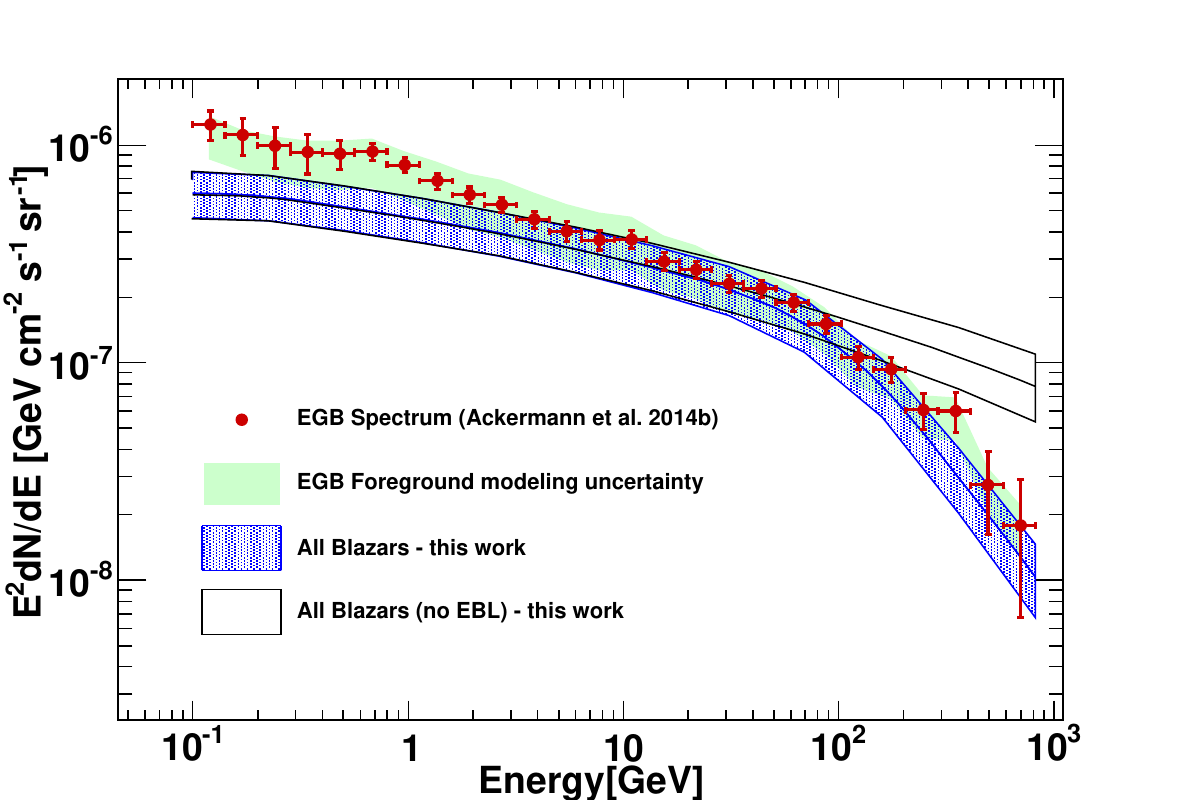}
\caption{EGB measured by the {\it Fermi}-LAT and the blazar contribution modeled through the $\gamma$-ray luminosity function~\cite{Ajello:2015mfa}.}
\label{gammaback}
\end{center}
\end{figure}

An analysis of the {\it Fermi}-LAT data indicates the importance of AGN that are {\it Fermi} sources in explaining the IGRB. Figure~\ref{blazarSC} 
shows their source count distribution~\cite{Stecker:2010di}. Figure~\ref{gammalum} shows the luminosity density of $\gamma$-rays from blazars as a function of redshift~\cite{Ajello:2013lka}. Figure~\ref{gammaback} shows the modeled contribution of {\it Fermi} blazars to the IGRB spectrum~\cite{Ajello:2015mfa}.

The fact that the sub-TeV $\gamma$-ray background is dominated by blazars tempts one to speculate that the neutrino sky is also dominated by blazars.
Indeed, the all-sky neutrino intensity from blazars can be estimated to be~\cite{Murase:2014foa}
\begin{eqnarray}
E_\nu^2\Phi_\nu&\sim&{10}^{-8}~{\rm GeV}~{\rm cm}^{-2}~{\rm s}^{-1}~{\rm sr}^{-1}~\xi_{\rm cr,2}\mathcal{R}_{\rm cr,2.5}^{-1}(\xi_z/8)\nonumber\\
&\times&\left(\frac{{\rm min}[1,f_{p\gamma}]}{0.05}\right)L_{\rm rad,48.5}{\left(\frac{\rho}{{10}^{-11.5}~{\rm Mpc}^{-3}}\right)}.\,\,\,\,\,\,\,\,\,\,\,
\end{eqnarray}
This estimate is based on our expectation that the all-sky neutrino intensity is typically dominated by FSRQs for a fixed value of $\xi_{\rm cr}$. Thus, in terms of energetics, it is possible for blazars to achieve the observed level of the all-sky neutrino intensity. 
Historically, the blazar contribution to the all-sky neutrino intensity was discussed in the hadronic scenario that explains the EGB measured by EGRET or $\gamma$-ray SEDs of the EGRET blazars~\cite{Mannheim:1998wp}. The fact that the IceCube neutrino flux is lower than the EGRET EGB flux disfavors the original model predictions because of $L_\nu\sim L_\gamma$. 
In lepto-hadronic models where $\gamma$-rays originate from nonthermal electrons, we may expect $L_\nu\lesssim L_\gamma$, but recent analyses have indicated that the blazar contribution is unlikely to be dominant in either leptonic or hadronic scenario. There are three relevant constraints. 

%
%

%

%
%

\begin{figure}[t]
\begin{center}
\includegraphics[width=0.49\linewidth]{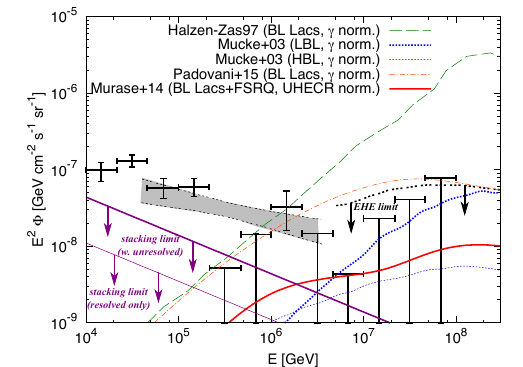}
\includegraphics[width=0.49\linewidth]{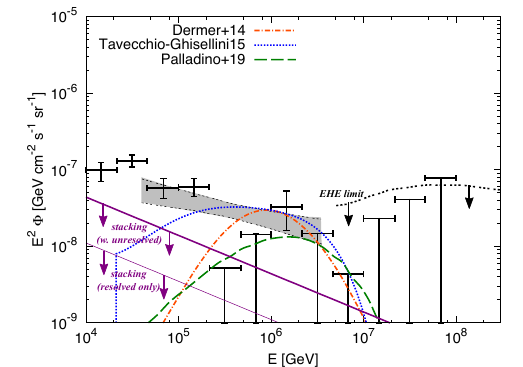}
\caption{(Left:) Comparison of some blazar models that are invoked to explain $\gamma$-ray and UHECR data~\cite{Halzen:1997hw,Muecke:2002bi,Murase:2014foa,Padovani:2015mba}. 
Earlier models before the IceCube discovery have been constrained by the EHE limit~\cite{IceCube:2018fhm}. 
(Right:) Comparison of some blazar models that are invoked to explain the IceCube data~\cite{Dermer:2014vaa,Tavecchio:2014xha,Palladino:2018lov}. The models to explain $\sim100$\% of the all-sky IceCube flux have been constrained by stacking limits~\cite{Hooper:2018wyk,Yuan:2019ucv}. 
}
\label{fig:blazarmodels}
\end{center}
\end{figure}

The first constraint comes from the EHE limit placed by the IceCube Collaboration~\cite{IceCube:2018fhm}. As shown in Figure~\ref{fig:blazarmodels}, the IceCube data have excluded some of the early model predictions, especially models that explain the EGB with proton-induced cascade emission. 
This supports that the $\gamma$-rays from at least some of the blazars should be leptonic in origin. Also, as previously mentioned, blazar models predict hard neutrino spectra. This is unavoidable especially below the PeV energy range. Thus, it is not easy for the blazar models to account for the neutrino data in the $10-100$~TeV range. 

It is possible for the blazars to account for the IceCube data above the 100~TeV energy. However, if the cosmic-ray spectrum is extended to ultrahigh energies, the models become incompatible with the IceCube data, as shown in Figure~\ref{fig:blazarmodels} left. 
Alternatively, it is possible to introduce a cutoff in the cosmic-ray spectrum. Some models are shown in Figure~\ref{fig:blazarmodels} right. Blazars are not the sources of UHECRs in these models, and lower energies of the spectral peaks could be explained by the stochastic acceleration mechanism~\cite{Dermer:2014vaa}. 
 
\begin{figure}[t]
\begin{center}
\includegraphics[width=0.7\linewidth]{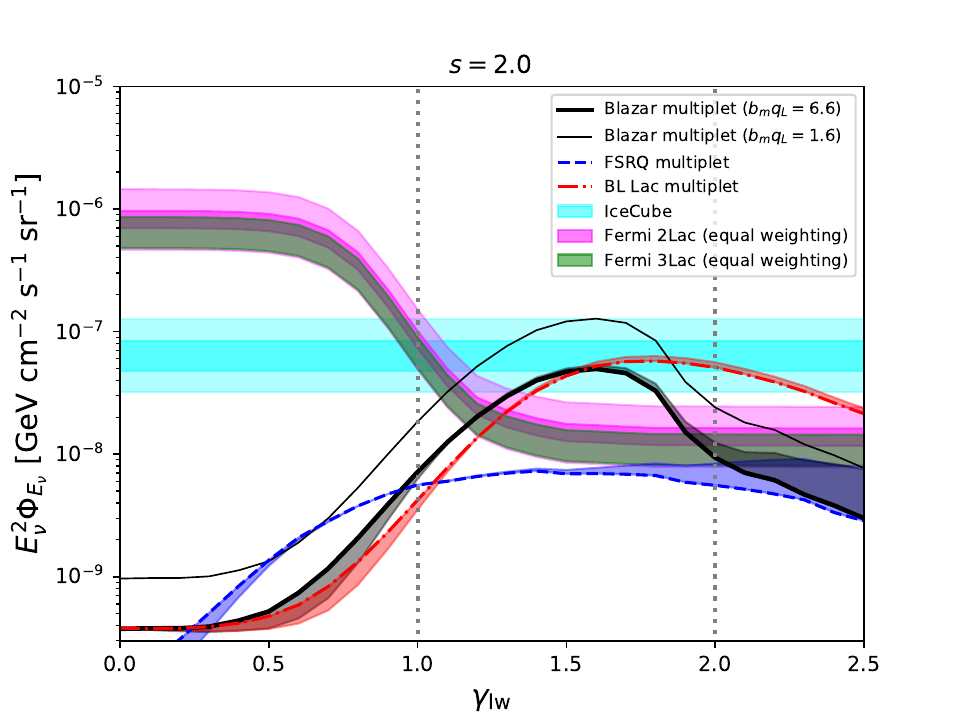}
\caption{Combination of stacking and $\gamma$-ray limits on the blazar contribution to the all-sky neutrino intensity as a function of the luminosity weighting index $\gamma_{\rm Lw}$. Adapted from Yuan et al.~\cite{Yuan:2019ucv}.
}
\label{fig:blazarlimit}
\end{center}
\end{figure}

The second constraint comes from stacking limits. The IceCube Collaboration reported the limits on the contribution from $\gamma$-ray detected blazars in the 2 year {\it Fermi}-LAT AGN Catalog (2LAC)~\cite{Aartsen:2016lir,Aartsen:2016oji}. Smith et al.~\cite{Smith:2020oac} presented the updated results with the 3LAC and 4LAC data, respectively, concluding that blazars account for at most 15\% of the neutrino background. All the results indicate that blazars responsible for the most of the $\gamma$-ray background account for only a small fraction of the neutrino background. However, these analyses have a caveat that possible contributions from uncatalogued blazars are not included~\cite{Palladino:2018lov}. Yuan et al.~\cite{Yuan:2019ucv} evaluated the contribution from all blazars including unresolved ones, using the $\gamma$-ray luminosity function obtained from the {\it Fermi}-LAT data, and concluded that the blazar contribution is subdominant for $\gamma_{\rm Lw}\gtrsim1$, which is consistent with typical predictions of the leptonic scenario that leads to $\gamma_{\rm Lw}\sim1.5-2$ (see Figure~\ref{fig:blazarlimit}.)   

The third constraint comes from the absence of multiplet neutrino sources in the neutrino data~\cite{Ahlers:2014ioa,Murase:2016gly}. The absence of multiple event sources leads to the following condition on the effective local number density of the sources $\rho^{\rm eff}$,
\begin{equation}
\hat{N}_s=b_{m,L}\left(\frac{\Delta\Omega}{3}\right)\rho^{\rm eff}d_{\rm lim}^3<1,
\label{eq:Nm_general}
\end{equation}
where $b_{m,L}$ is an order-of-unity factor that depends on details of analyses, $\Delta \Omega$ is the solid angle covered by the detector, and $d_{\rm lim}$ is given by
\begin{equation}
d_{\rm lim}={\left(\frac{\varepsilon_\nu L_{\varepsilon_{\nu_\mu}}}{4\pi F_{\rm lim}}\right)}^{1/2}.
\end{equation}
Here $\varepsilon_\nu L_{\varepsilon_{\nu_\mu}}$ is the differential muon neutrino luminosity and $F_{\rm lim}$ is the flux sensitivity. For example, if the number of false multiplet sources is negligible, $N_b\lesssim1$, we obtain $b_{m,L}\simeq6.6q_L$ for $m\geq2$ multiplets. In the background dominated limit, we may use $b_{m,L}=1$. 
For an $E_{\nu}^{-2}$ neutrino spectrum, the local number density of the sources is constrained as~\cite{Murase:2018iyl}
\begin{eqnarray}\label{eq:n0a}
\rho^{\rm eff}&\lesssim&1.9\times10^{-10}~{\rm Mpc^{-3}}~\left(\frac{\varepsilon_\nu L_{\varepsilon_{\nu_\mu}}}{{10}^{44}~{\rm erg\,s^{-1}}}\right)^{-3/2}{\left(\frac{b_mq_L}{6.6}\right)}^{-1}F_{\rm lim,-9.2}^{3/2}\left(\frac{2\pi}{\Delta\Omega}\right),\,\,\,
\end{eqnarray}
where $q_L\sim1-3$ is a luminosity-dependent correction factor determined by the redshift evolution, and the 8 year point-source sensitivity (90\% CL), $F_{\rm lim}\sim(5-6)\times{10}^{-10}~{\rm GeV cm^{-2}s^{-1}}$, is used as a reference~\citep{Aartsen:2017kru}.
Then, the contribution to the all-sky neutrino intensity is constrained as
\begin{eqnarray}
E_\nu^2\Phi_\nu
&\lesssim&6.9\times{10}^{-9}~{\rm GeV}~{\rm cm}^{-2}~{\rm s}^{-1}~{\rm sr}^{-1}~\left(\frac{\xi_z}{0.7}\right){\left(\frac{6.6}{b_mq_L}\right)}^{2/3}\nonumber\\
&\times&\left(\frac{\rho^{\rm eff}}{{10}^{-7}~{\rm Mpc}^{-3}}\right)^{1/3}F_{\rm lim,-9.2}{\left(\frac{2\pi}{\Delta\Omega}\right)}^{2/3}.
\label{eq:diffuse2}
\end{eqnarray}

%
%

%
%

We note that the second and third constraints are complementary in a sense. The stacking limits can be significant relaxed if $\gamma$-ray dim blazars are dominant in the neutrino sky, i.e., $\gamma_{\rm Lw}\lesssim1$. However, such blazars are mostly BL Lac objects, for which the multiplet constraints are more important. As a result, the three constraints suggest that the all-sky neutrino intensity cannot be solely explained by blazars, and their contribution is $\lesssim30$\%~\cite{Yuan:2019ucv}. This conclusion is especially the case for the origin of medium-energy neutrinos in the $10-100$~TeV range. 
However, it is still possible for them to give a significant contribution especially in the PeV range. 
This is especially the case if high-redshift FSRQs whose $\gamma$-ray peak is located in the MeV range make a significant contribution. Such MeV blazars exist at large redshifts with $z\gtrsim2-4$, in which the multiplet constraints are significantly weakened as pointed by Murase and Waxman~\cite{Murase:2016gly}. A large fraction of the MeV blazars may also be missed in the {\it Fermi} catalogues.  
Along this line, the recent claim of a significant correlation between radio-selected AGN and muon neutrinos may be intriguing~\cite{Plavin:2020emb}. Radio-bright AGN in the sample are mostly FSRQs, which are considered as efficient neutrino emitters. However, the existence of the correlation has also been questioned~\cite{Zhou:2021rhl}, and the significant correlation does not necessarily mean that the blazars are dominant in the neutrino sky.  

Finally, we remark on the connection between UHECRs and neutrinos. UHECRs may be accelerated in inner jets of jet-loudAGN~\cite{Murase:2011cy}. If the cosmic-ray spectrum is extended to ultrahigh energies, the neutrino spectra are typically expected in the PeV-EeV range, which is especially the case if the cosmic-ray spectrum is as hard as $s_{\rm cr}\sim2$. Murase et al.~\cite{Murase:2014foa} found that the resulting neutrino emission is likely to be dominated by QHBs, where the photomeson production with external radiation fields play crucial roles and UHECR nuclei are efficiently depleted. They also suggested the blazar-UHECR scenario, in which UHECR nuclei are dominated by BL Lac objects because the nucleus-survival condition is easily satisfied in such low-luminosity objects~\cite{Murase:2011cy}. The UHECR nuclei from the blazar jets are likely to be deflected by structured magnetic fields as well as radio lobes or cocoons, while high-energy neutrinos are beamed to the on-axis observer. Blazars, especially FSRQs, are promising EeV neutrino emitters, whose all-sky flux can overwhelm the cosmogenic neutrino flux, being excellent targets for next-generation ultrahigh-energy neutrino detectors. These findings are also supported by more recent independent studies~\cite{Righi:2020ufi,Rodrigues:2020pli}. 

However, one should keep in mind that jet-loudAGN do not have to be strong neutrino emitters in more general. Efficient neutrino production is expected in inner jets, while the UHECR acceleration is often considered at large-scale jets in the galaxy scale~\cite{Dermer:2008cy,Kimura:2017ubz,Matthews:2018rpe}. 
In these UHECR acceleration models, the expected neutrino flux in the EeV range from the sources is much lower and the cosmogenic neutrino flux can be more important~\cite{BeckerTjus:2014uyv,Zhang:2018agl}, although nearby objects such as Cen A could still be detectable~\cite{Cuoco:2007aa,Koers:2008hv,Kachelriess:2008qx}.

\subsection{TXS 0506+056 and Other Blazar Coincidences}
In 2017, a high-energy neutrino event, IceCube-170922A with an energy of $E_\nu\sim0.3$~PeV, was detected in IceCube~\cite{Aartsen2018blazar1}. Follow-up observations have been made over the world. With help of a list of blazars observed by the Kanata telescope, {\it Fermi} sources were searched, and the $\gamma$-ray counterpart was identified, which was the blazar, TXS 0506+0656 at $z=0.336$. Interestingly, this blazar had also been detected as one of the EGRET sources~\cite{2001AIPC..587..251D}. Independently, X-ray counterparts the {\it Swift} follow-up observation was triggered, and TXS 0506+056 was found as one of the promising counterparts showing the active state. This blazar was also observed as a flaring blazar by {\it NuSTAR}, MAGIC~\cite{MAGIC:2018sak}, and other radio facilities. It is rare to find such a flaring blazar in the random sky, and the significance is estimated to be $\sim3\sigma$.  

In the subsequent analysis, the IceCube Collaboration searched for neutrino emission in the archival data of TXS 0506+056, and they found a $\sim3.5\sigma$ excess, corresponding to $13\pm5$ excess events, in the 2014-2015 period~\cite{Aartsen2018blazar2}. However, any flaring activity was not found in both X-ray and $\gamma$-ray data, so the 2014-2015 flare is regarded as an orphan neutrino flare. The combination of this 2014-2015 neutrino flare and the 2017 multimessenger flare is intriguing even though it is too early to be conclusive about this blazar as a source of high-energy neutrinos.    

\begin{figure}[t]
\begin{center}
\includegraphics[width=0.7\linewidth]{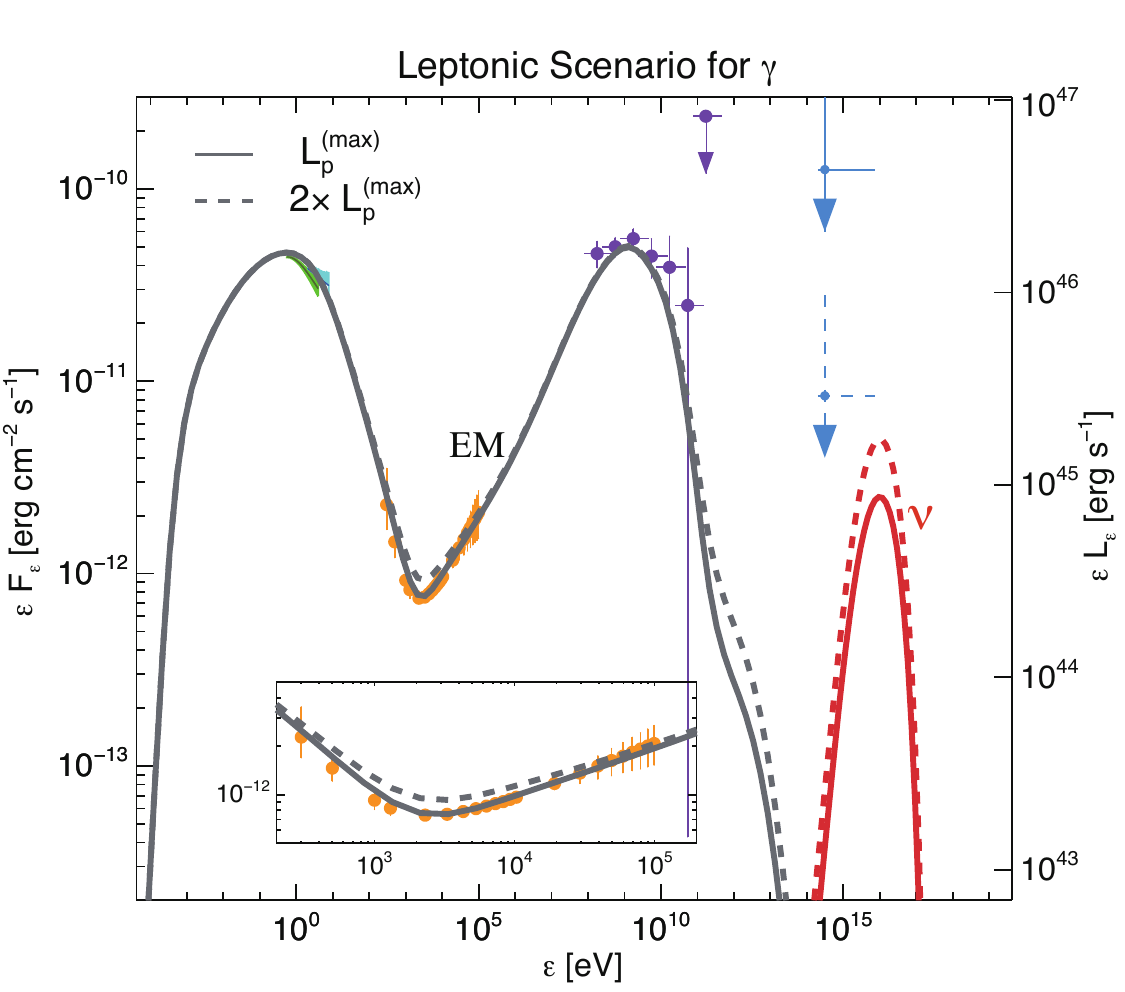}
\caption{Multimessenger SEDs of the 2017 multimessenger flare from TXS 0506+056, with theoretical curves by the single-zone lepto-hadronic modeling~\cite{Keivani:2018rnh}. The leptonic scenario for $\gamma$-rays is considered. Optical and ultraviolet data are from Xshooter and {\it Swift}-UVOT, X-ray data are from {\it Swift}-XRT and {\it NuSTAR}, and $\gamma$-ray data are from {\it Fermi}-LAT. The neutrino data by IceCube are also shown for different durations, 0.5~years (upper) and 7.5~years (lower).     
}
\label{fig:txssed}
\end{center}
\end{figure}

\begin{figure}[t]
\begin{center}
\includegraphics[width=0.7\linewidth]{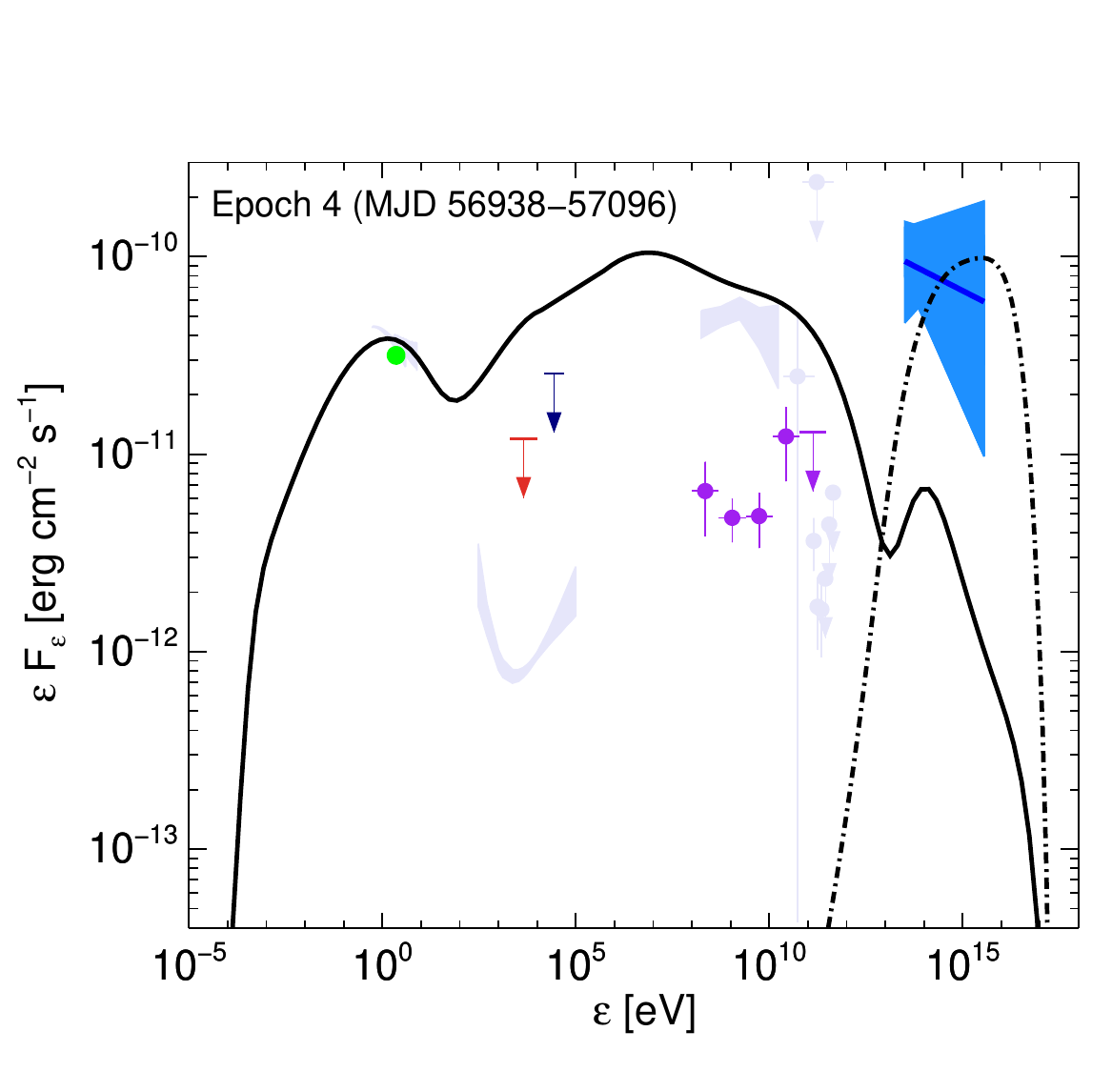}
\caption{Multimessenger SEDs of the 2014-2015 neutrino flare from TXS 0506+056, with theoretical curves by the single-zone lepto-hadronic modeling~\cite{Petropoulou:2019zqp}. Optical data are from ASAS-SN, X-ray data are from {\it MAXI} and {\it Swift}-XRT, and $\gamma$-ray data are from {\it Fermi}-LAT. The neutrino data by IceCube are also shown.  
}
\label{fig:txssed2}
\end{center}
\end{figure}

\begin{figure}[t]
\begin{center}
\includegraphics[width=0.8\linewidth]{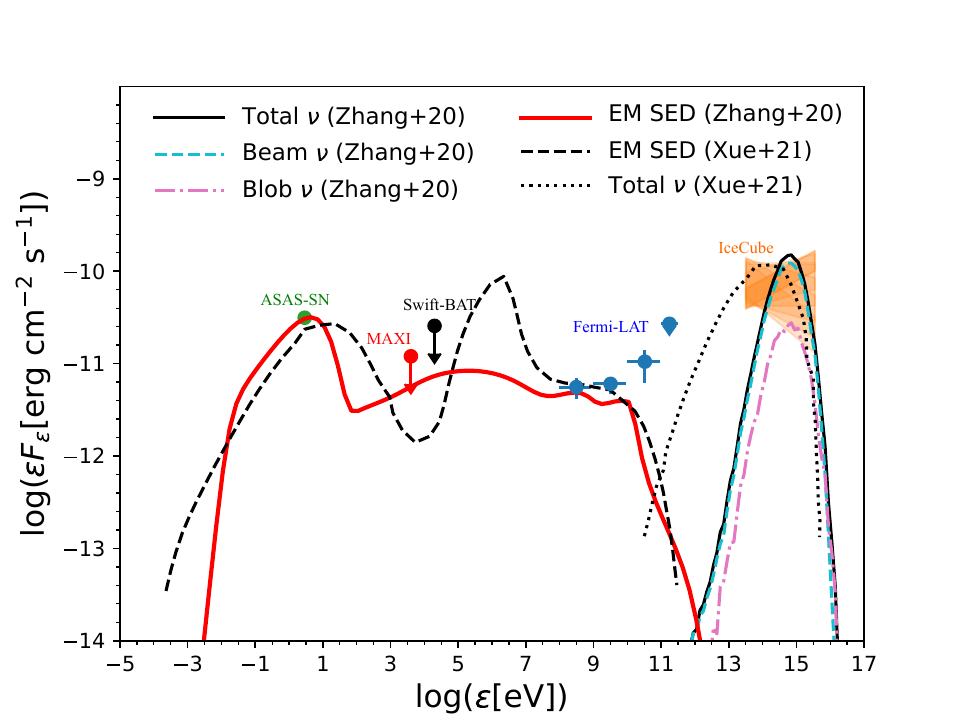}
\caption{Multimessenger SEDs of the 2014-2015 neutrino flare from TXS 0506+056, with theoretical curves of some multizone models. As demonstrative examples, the neutral beam model by Zhang et al.~\cite{Zhang:2019htg} and the two-zone model by Xue et al.~\cite{Xue:2019txw} are shown. They have different predictions for MeV $\gamma$-ray spectra. 
}
\label{fig:txssed3}
\end{center}
\end{figure}

Thanks to dedicated observational campaigns in 2017, the multiwavelength SED of TXS 0506+056 at this epoch were measured quite well, which are shown in Figure~\ref{fig:txssed}. Figures~\ref{fig:txssed} -~\ref{fig:txssed3} also sketch some theoretical model SEDs. The {\it Swift}-UVOT and Xshooter observations indicate that the peak frequency is located around $\sim3\times10^{14}$~Hz~\cite{Keivani:2018rnh}, which implies that TXS 0506+056 is classified as a low or intermediate synchrotron peak BL Lac object (see also References~\cite{Morokuma:2020xtl,Hwang:2020ycs}), although other observations imply that this blazar is a masquerading blazar~\cite{Padovani:2019xcv}.
The SED has been modeled by various groups, and it is possible to explain the multiwavelength spectrum both in leptonic and hadronic scenarios.  
In the leptonic scenario, detailed optical and ultraviolet data are crucial, and Keivani et al.~\cite{Keivani:2018rnh} found that it is difficult to explain the SED in the simplest synchrotron self-Compton scenario, and introduced an additional external radiation field that can be scattered accretion disk emission or photons from the sheath region of the AGN jet~\cite{Keivani:2018rnh,MAGIC:2018sak}.
In the hadronic scenario, $\gamma$-ray emission can be attributed to proton synchrotron radiation by UHECRs~\cite{Keivani:2018rnh,Gao:2018mnu,Cerruti:2018tmc}. 

%

%

However, the physical association between neutrinos and this blazar are challenged at least in the simplest single-zone modeling. In the case of the 2017 neutrino flare, only one neutrino event was detected, so the interpretation is subject to the large statistical fluctuation, which is sometimes called the Eddington bias~\cite{Strotjohann:2018ufz}. Nevertheless, the observation of IceCube-170922A for a given duration of 0.5 years indicates that the required neutrino luminosity is $L_\nu\sim10^{46}-10^{47}~{\rm erg}~{\rm s}^{-1}$. High-energy neutrinos originate from charged pions, and high-energy $\gamma$-rays must be accompanied by electromagnetic cascades due to the neutral pion production and the Bethe-Heitler pair production process. The resulting cascade flux generally has a very broad spectrum, which is constrained by the observed beautiful X-ray valley seen by {\it Swift} and {\it NuSTAR}. As a result, the neutrino luminosity in the $0.1-1$~PeV range is constrained as $L_\nu\lesssim{10}^{44}-10^{45}~{\rm erg}~{\rm s}^{-1}$. In other words, even in the best case scenario, the expected number of neutrinos in realtime observations is at most $\sim0.01$, which requires a large Poisson fluctuation to account for the neutrino observation~\cite{Keivani:2018rnh}. 
The situation is more serious for the 2014-2015 flare. The excess neutrino emission indicates that the neutrino luminosity is $L_\nu\sim10^{46}-10^{47}~{\rm erg}~{\rm s}^{-1}$, comparable to that of the 2017 multimessenger flare. However, archival data of {\it MAXI}, {\it Swift}, and {\it Fermi} suggest that there is no flaring activity in this period, and the X-ray upper limits and $\gamma$-ray data are in strong tension with the cascade flux resulting from the observed neutrino flux~\cite{Murase:2018iyl,Rodrigues:2018tku,Reimer:2018vvw,Petropoulou:2019zqp,Gasparyan:2021oad}. 

These cascade constraints are quite robust and insensitive to details of the models. They basically rely on the energy conservation, and the argument largely holds because electromagnetic cascades lead to broad spectra in the X-ray and $\gamma$-ray range.
For X-ray emission of TXS 0506+056, synchrotron emission from pairs injected via the Bethe-Heitler process is important~\citep{Keivani:2018rnh}, and the minimum synchrotron cascade flux associated with the neutrino flux at $\varepsilon_\nu$ is estimated to be~\cite{Murase:2018iyl} 
\begin{eqnarray}
\varepsilon_{\gamma}L_{\varepsilon_\gamma}|_{\varepsilon_{\rm syn}^{\rm BH}}\approx\frac{1}{2(1+Y_{\rm IC})}g[\beta]f_{p\gamma}(\varepsilon_{p}L_{\varepsilon_p})
\approx\frac{4g[\beta]}{3(1+Y_{\rm IC})}\varepsilon_{\nu}L_{\varepsilon_\nu},
\label{eq:BHsyn}
\end{eqnarray}
where $\beta$ is the photon index that is $\beta=2.8$ for the 2017 multimessenger flare of TXS 0506+056, $g[\beta]\sim0.011{(30)}^{\beta-1}$, $\varepsilon_{\rm syn}^{\rm BH}\sim 6~{\rm keV}~{B'}_{-0.5}{(\varepsilon_p/6~{\rm PeV})}^2(20/\delta)$ is the characteristic frequency of synchrotron emission by pairs from protons with $\varepsilon_p\sim20\varepsilon_\nu$, $B'$ is the comoving magnetic field strength, and $Y_{\rm IC}$ is the possible inverse Compton Y parameter. 
Similarly, for synchrotron emission from pairs injected via the photomeson production and two-photon annihilation for pionic $\gamma$-rays, the synchrotron cascade flux is,
\begin{eqnarray}
\varepsilon_{\gamma}L_{\varepsilon_\gamma}|_{\varepsilon_{\rm syn}^{p\gamma}} 
\approx \frac{1}{2(1+Y_{\rm IC})}\frac{5}{8}f_{p\gamma}(\varepsilon_{p}L_{\varepsilon_p})
\approx \frac{5}{6(1+Y_{\rm IC})}\varepsilon_{\nu}L_{\varepsilon_\nu},
\label{eq:pgsyn}
\end{eqnarray}
where $\varepsilon_{\rm syn}^{p\gamma}\sim60~{\rm MeV}~{B'}_{-0.5}{(\varepsilon_p/6~{\rm PeV})}^2(20/\delta)$. Equations (\ref{eq:BHsyn}) and (\ref{eq:pgsyn}) clearly indicate that the synchrotron cascade flux, which is expected in the X-ray and $\gamma$-ray range, is comparable to the neutrino flux.  
We also note that the similar issue exists even if neutrinos originate from $pp$ interactions rather than $p\gamma$ interactions~\cite{Murase:2018iyl}. We do not discuss hadronuclear neutrino production models because these models usually suffer from other issues such as the overburdened jet and cosmic-ray isotropization~\cite{Atoyan:2002gu}.   

In addition to the above cascade constraints, there are other difficulties in interpreting the multimessenger data. For the 2017 flare, the MAGIC collaboration detected very high-energy $\gamma$-rays~\cite{MAGIC:2018sak}, which implies that they must escape from the source given that neutrinos and $\gamma$-rays are produced in the same region. Then, imposing $\tau_{\gamma\gamma}<1$ at 100~GeV leads to
\begin{equation}
f_{p\gamma}<{10}^{-3}~{(\varepsilon_p/60~{\rm PeV})}^{\beta-1},
\end{equation}
which implies that the photomeson production efficiency is rather low. In other words, the isotropic-equivalent luminosity of cosmic rays is constrained as $L_p\gtrsim 10^{50}~{\rm erg}~{\rm s}^{-1}$ and the nonthermal proton-to-electron ratio is as large as $L_p/L_e \gtrsim 300$~\cite{Keivani:2018rnh,Murase:2018iyl}. The former means that the the jet luminosity is significantly larger than the Eddington luminosity for a typical jet opening angle of $\sim0.1$~rad. Large values of $L_p/L_e$ would also be challenging for particle acceleration theories especially if the particles are accelerated by magnetic reconnections (see also Reference~\cite{Zdziarski:2015rsa}). In either case, our understanding of blazar emission and underlying jet physics would need a significant revision if TXS 0506+06 is confirmed as a real neutrino source.   

We also note that TXS 0506+056 would not be an UHECR accelerator given the physical association with IceCube-170922A. If the cosmic-ray spectrum is extended to ultrahigh energies with $s_{\rm cr}\sim2$, the predicted neutrino spectrum is so hard that it would contradict the nondetection of $>10$~PeV neutrinos during the flare. This rules out the hadronic model accounting for the observed $\gamma$-ray data, as a scenario that simultaneously explains the detection of IceCube-170922A~\cite{Keivani:2018rnh,Gao:2018mnu}. On the other hand, the leptonic scenario is viable, as shown in Figure~\ref{fig:txssed}, but the proton spectrum cuts off at $10-100$~PeV~\citep{Keivani:2018rnh}.

The difficulty of the single-zone modeling motivates the development of multizone models. Considering different emission regions is natural in view of the degree of complexity seen in radio emission regions. 
As one of possibilities, two-zone models have been considered, where neutrinos mainly come from the inner dissipation region and $\gamma$-rays originate from the outer dissipation region~\cite{Rodrigues:2018tku,Xue:2019txw}. However, the number of model parameters is basically doubled, so it is not easy to make testable predictions. Murase et al.~\cite{Murase:2018iyl} and Zhang et al.~\cite{Zhang:2019htg} applied a neutral beam model~\cite{Atoyan:2002gu,Dermer:2012rg} to explain the multimessenger data. This model is an extension of single-zone models, in which escaping cosmic rays that are presumably neutrons enhance the flux of neutrinos. A nice feature of this model is that the beam-induced cascade $\gamma$-ray flux is suppressed due to the deflection and time delay through magnetic fields.   

Another puzzle brought by TXS 0506+056 is why this intermediate luminosity blazar was found to be the brightest neutrino source among many more brighter blazars in photons. As one of the possibilities, it has been speculated that this blazar forms a SMBH binary based on the radio data~\cite{Kun:2018zin,Britzen:2019nfc}. However, the interpretation is still under debate because the radio data are also consistent with a structured jet~\cite{Ros:2019bgo,Li:2020wby,Sumida:2021uov}. Note that the cascade constraint is applied anyway whether TXS 0506+056 is an atypical blazar or not.   

Because we currently lack a convincing, concordance picture of TXS 0506+056, it is crucial to search for more coincidences between neutrinos and blazars. Along this line, Oikonomou et al.~\cite{Oikonomou:2019djc} investigated the detectability of blazar flares with current and future neutrino detectors. The detection may be challenging if one assumes moderate cosmic-ray loading factors motivated by the blazar-UHECR hypothesis, but can be promising if a lot of cosmic rays are loaded in the jet as suggested by TXS 0506+056. Intriguingly, additional hints were also reported.  

3HSP J095507.9+355101 is an extreme blazar, which may be associated with a high-energy neutrino event, IceCube-200107A~\cite{Giommi:2020viy}. This blazar showed a hard X-ray flare around the detection of the neutrino event. However, the Poisson probability to detect a single muon neutrino with the effective area for realtime alerts is as low as $\sim1$\% even if 10 years of IceCube observations are assumed~\cite{Petropoulou:2020pqh}.  

PKS 1502+106 is a FSRQ that may be associated with IceCube-190730A. Contrary to 3HSP J095507.9+355101, this blazar was in a quiet state at the time of the neutrino alert. If IceCube-190730A is attributed to steady emission from this blazar, the required cosmic-ray luminosity can be consistent with the value required for blazars to explain the observed UHECR flux, which is also below the Eddington luminosity~\cite{Rodrigues:2020fbu,Oikonomou:2021akf}. 

\section{AGN Embedded in Magnetized Environments}
\label{sec:environment}
Neutrinos can be produced in magnetized environments surrounding AGN.
Cosmic rays that are accelerated in the acceleration region eventually escape and can be confined in the environments for a long time. 
Neutrino and $\gamma$-ray production naturally occur during their confinement time, and promising ``cosmic-ray reservoir'' sources being galaxy clusters and groups and starburst galaxies~\cite{Murase:2013rfa}. 

%
Galaxy clusters and groups are of particular interest in light of the multimessenger connection among the observed all-sky astroparticle intensities~\cite{Berezinsky:1996wx,Colafrancesco:1998us}. Jet-loud AGN are believed to be the most promising sites for the production of UHECRs, where not only inner jets but also large-scale jets have been discussed. Viable ion acceleration mechanisms include the one-shot shear acceleration~\cite{Caprioli:2015zka,Kimura:2017ubz,Mbarek:2021bay} and shock acceleration in backflows resulting from jet-medium interactions~\cite{Matthews:2018rpe,Bell:2019nnf}. 
In addition to jet-loudAGN, weak jets of RQ AGN may accelerate cosmic rays to ultrahigh energies~\cite{Pe'er:2009rc}, and AGN winds have also been suggested as possible cosmic-ray accelerators~\cite{Tamborra:2014xia,Wang:2016oid,Liu:2017bjr}. 
Low-energy cosmic rays are likely be confined in a cocoon and subject to energy losses during the cocoon expansion, and only sufficiently high-energy cosmic rays, including UHECRs, would escape without significant adiabatic losses. The escaping high-energy cosmic rays can further be confined in intra-cluster material for a cosmological time.  

\begin{figure}[t]
\begin{center}
\includegraphics[width=0.9\linewidth]{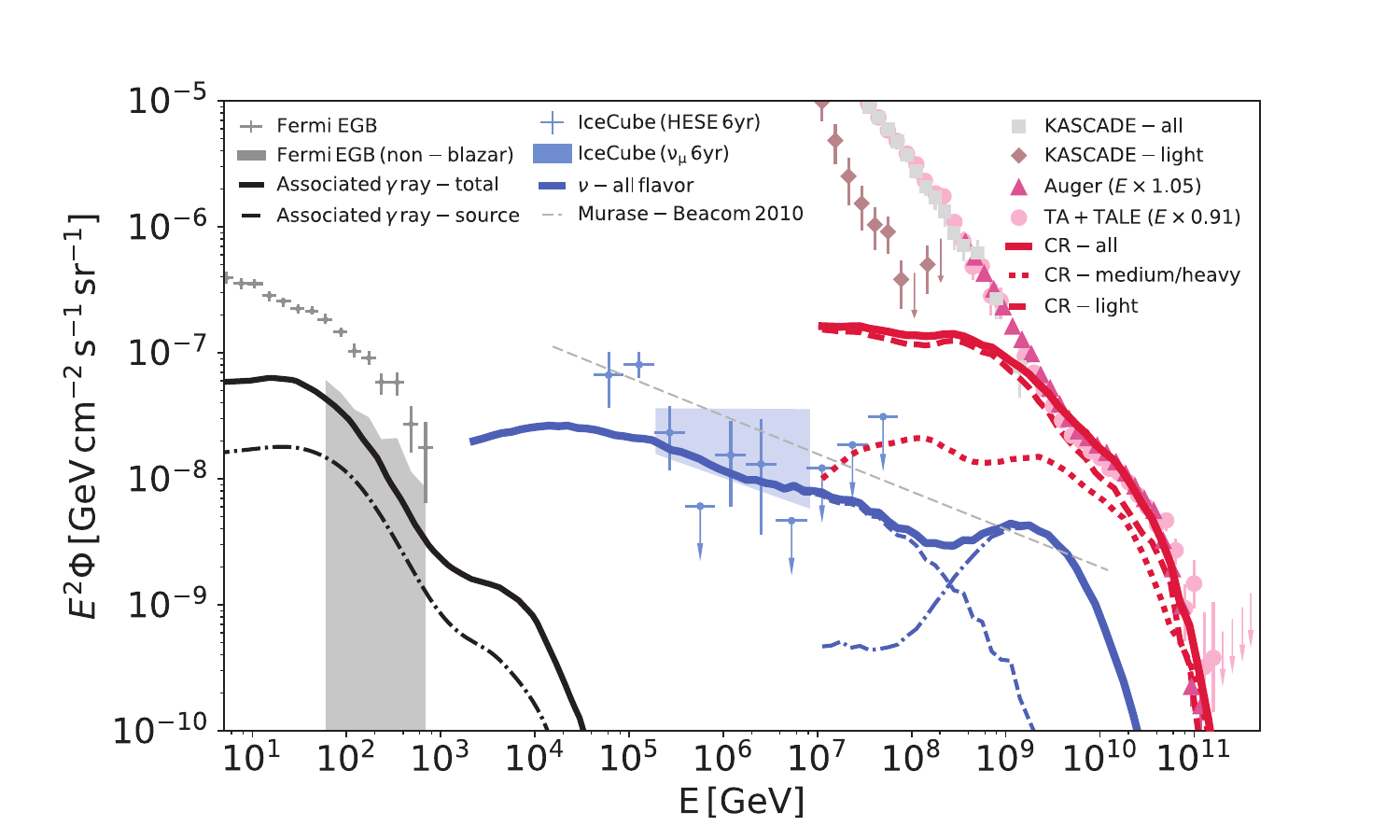}
\caption{Predictions of high-energy neutrino, $\gamma$-ray, and UHECR spectra in the astroparticle grand-unification scenario, in which AGN embedded in galaxy clusters and groups are considered. Adapted from Fang and Murase~\cite{Fang:2017zjf}. 
}
\label{fig:unified}
\end{center}
\end{figure}

The effective optical depth for inelastic $pp$ collisions is estimated to be~\cite{Murase:2008yt,Murase:2013rfa} 
\begin{equation}
f_{pp}\simeq1.1\times{10}^{-2}~g\bar{n}_{-4}(t_{\rm esc}/3~{\rm Gyr}),
\end{equation}
where $n$ is the intra-cluster gas density, $g$ is the possible enhancement factor due to the cluster or group density profile, and $t_{\rm esc}$ is the cosmic-ray escape time. Assuming that accretion shocks of galaxy clusters or AGN jets are the sources of UHECRs or sub-ankle cosmic rays above the second knee, Murase et al.~\cite{Murase:2008yt} and Kotera et al.~\cite{Kotera:2009ms} predicted that the all-sky neutrino intensity is around $E_\nu^2\Phi_\nu\sim{10}^{-9}-{10}^{-8}~{\rm GeV}~{\rm cm}^{-2}~{\rm s}^{-1}~{\rm sr}^{-1}$, which is consistent with the IceCube data above 100~TeV energies. 
By extending these cosmic-ray reservoir models, Fang and Murase~\cite{Fang:2017zjf} proposed that the all-sky fluxes of three messenger particles can be explained in a unified manner (see Figure~\ref{fig:unified}). In this ``astroparticle grand-unification'' scenario, sub-PeV neutrinos are mainly produced via $pp$ interactions inside the intra-cluster medium. UHECRs escape and contribute to the observed UHECR flux, and spectra of the UHECRs injected into intergalactic space are hard because of the magnetic confinement and nuclear photodisintegration in galaxy clusters. Cosmogenic $\gamma$-rays and cascade emission induced by $\gamma$-rays generated inside galaxy clusters and groups contribute the IGRB especially in the sub-TeV range. 
This model generally predicts the smooth transition from source neutrinos to cosmogenic neutrinos that are dominant in the EeV range, which will be testable with next-generation neutrino telescopes such as IceCube-Gen2~\cite{IceCube-Gen2:2020qha}, GRAND~\cite{GRAND:2018iaj}, and Trinity~\cite{Otte:2019knb}. 

%
%

%

%
%

\section{Tidal Disruption Events}\label{sec:tidal}
A TDE occurs when a star in orbit around a SMBH gets close enough to the SMBH to be disrupted by the its tidal force. Then about half of the the stellar debris falls back to the SMBH presumably presumably at a super-Eddington rate, while the other half is launched as an outflowing debris. The former would involve a characteristic flare that can last for months to years~\cite{Rees:1988bf}. 

UHECR production in TDEs was suggested by Farrar and Gruzinov as a giant AGN flare model~\cite{Farrar:2008ex}, and the resulting neutrino emission was calculated following this scenario~\cite{Murase:2008zzc}. After the {\it Swift} discovery of a jetted TDE~\cite{Burrows:2011dn,Bloom:2011xk}, high-energy neutrino production in jetted TDEs has been investigated in detail~\cite{Wang:2011ip,Wang:2015mmh,Dai:2016gtz,Senno:2016bso,Lunardini:2016xwi}. Jetted TDE models have also been of interest in light of the possible connection to UHECRs~\cite{Zhang:2017hom,Biehl:2017hnb,Guepin:2017abw,Yoshida:2020div}. 

\begin{figure}[t]
\begin{center}
\includegraphics[width=0.6\linewidth]{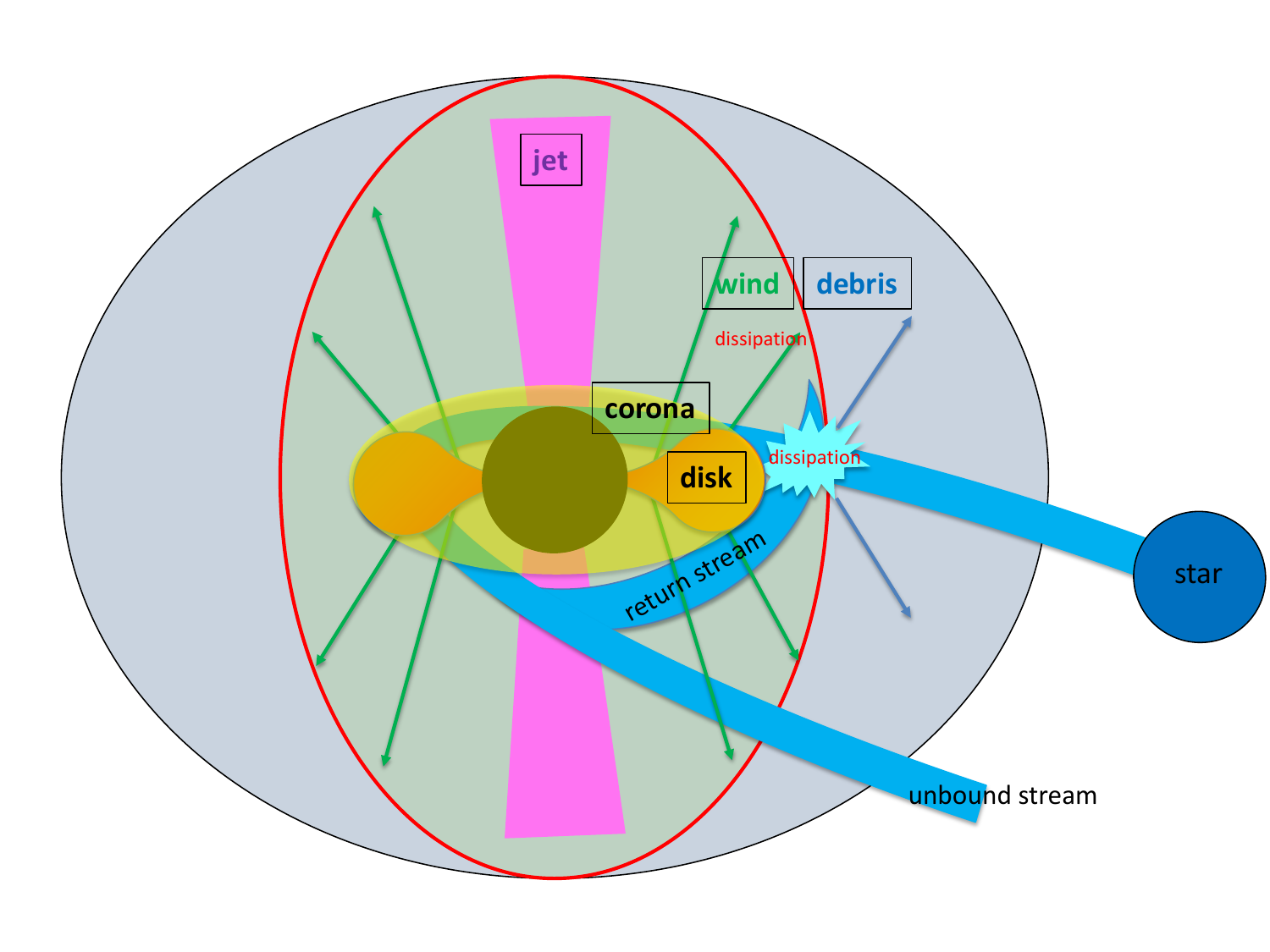}
\caption{Schematic picture of high-energy neutrino production in TDEs. Cosmic rays that can be accelerated at different acceleration sites, including the vicinity of a SMBH, shocks induced by sub-relativistic winds or tidal streams, and relativistic jets. Adapted from Murase et al.~\cite{Murase:2020lnu}.}  
\label{fig:TDE}
\end{center}
\end{figure}

Only a fraction of TDEs have powerful jets, and most of them are non-jetted and observed as optical and ultraviolet transients. 
Recently, Stein et al.~\cite{Stein:2020xhk} reported that one of such non-jetted TDEs, AT 2019dsg, was associated with the recent detection of a $\sim200$~TeV neutrino, IceCube-191001A, by the IceCube collaboration. The radio observations strongly constrain the jet component~\cite{Cendes:2021bvp,Matsumoto:2021qqo}, and high-energy neutrino production models involving accretion disks and their coronae (see Figure~\ref{fig:TDE}) have been considered~\cite{Hayasaki:2019kjy,Murase:2020lnu}. More data are necessary to confirm whether this association is physical or not, but it may indicate that AGN and TDEs produce high-energy neutrinos through the similar mechanism.

In May of 2022 another candidate TDE was reported by Reusch et al. to be coincident with a high energy neutrino, viz., AT2019fdr~\cite{Reusch:2021ztx}. 
The source object was an AGN similar to the TDE associated The host galaxy was an AGN, while AT2019dsg is associated with a galaxy in the green valley. 
Based on their statistical analysis, Reusch et al. argued that about 10\% of high-energy astrophysical neutrinos might come from TDEs~\cite{Reusch:2021ztx}. 
These observations may indicate that AGN and TDEs produce high-energy neutrinos via a similar production mechanism. More data are necessary to confirm these conclusions.

\section{Summary and Prospects}\label{sec:summary}
We can summarize the basic points regarding neutrino production in AGN discussed in this chapter as follows:
\begin{itemize}
    \item The vicinity of SMBHs is the promising site for efficient neutrino production. Given that ion acceleration occurs, cosmic rays are efficiently depleted through $p\gamma$ and $pp$ interactions. Detecting neutrinos from such compact regions of AGN will give us new insight into plasma dissipation and particle energization in dense environments. The most significant source in the IceCube point source analysis, NGC 1068, has also been theoretically expected to be the most promising neutrino source in the IceCube sky, and various models can be test by upcoming multimessenger observations.
    
    \item Inner jets are known to be the site of high-energy $\gamma$-rays, so it is natural to expect that high-energy neutrino production occurs as well. On-axis objects, i.e., blazars, are promsing neutrino sources especially in the PeV-EeV range. Detecting neutrinos from blazars will give us crucial clues to not only the origin of UHECRs but also the physics of relativistic jets, particle acceleration and associated nonthermal radiation. 
    Associations of neutrinos with some blazars such as TXS 0506+056 are not yet understood, and further investigations are necessary both theoretically and observationally. 
    
    \item Large-scale jets of AGN are among the most promising sites for UHECR acceleration. High-energy neutrino production inside the acceleration zone may not be much efficient, but cosmic rays may further be confined in surrounding magnetized environments for a long time. Galaxy clusters and groups, as well as star-forming galaxies coexisting with AGN, serve as cosmic-ray reservoirs, which may significantly contribute to the observed neutrino flux. Such models also predict the strong neutrino--$\gamma$-ray connection, may provide a grand-unification picture of three messengers including cosmic rays.  
   
\end{itemize}

Our understanding of AGN physics has matured over the past decade through not only dedicated multiwavelength observations from radio to $\gamma$-ray bands. In addition, MHD and PIC simulations have also deepened our physical understanding of how black hole systems release gravitational energy, how their coronae are formed, how winds and jets are launched, how magnetic fields dissipate, and how particles are accelerated. 

Now, the golden era of multimessenger astrophysics has begun with many new questions. Future facilities using new detection techniques, as discussed in detail in Chapters 5, 6 and 7, hold great prospects for answering at least some of these questions. 

\bibliographystyle{ws-rv-van}
\bibliography{kmurase}
\end{document}